\documentclass[prd,aps,preprintnumbers,nofootinbib,superscriptaddress]{revtex4-1}
\usepackage{graphicx}
\usepackage{amssymb}
\usepackage{amsmath}
\usepackage{booktabs}
\usepackage{dcolumn}
\newcolumntype{x}[1]{D{.}{.}{#1}}

\newcommand{\beq}{\begin{eqnarray}}
\newcommand{\eeq}{\end{eqnarray}}

\def\fsl#1{\setbox0=\hbox{$#1$}           
   \dimen0=\wd0                                 
   \setbox1=\hbox{/} \dimen1=\wd1               
   \ifdim\dimen0>\dimen1                        
      \rlap{\hbox to \dimen0{\hfil/\hfil}}      
      #1                                        
   \else                                        
      \rlap{\hbox to \dimen1{\hfil$#1$\hfil}}   
      /                                         
   \fi}                                         %

\begin{document}

\preprint{KEK Preprint 2012-47}

\title{Walking signals in $N_f=8$ QCD on the lattice}

\author{Yasumichi~Aoki}
\affiliation{Kobayashi-Maskawa Institute for the Origin of Particles and the Universe, \\
Nagoya University, Nagoya 464-8602, Japan}

\author{Tatsumi~Aoyama}
\affiliation{Kobayashi-Maskawa Institute for the Origin of Particles and the Universe, \\
Nagoya University, Nagoya 464-8602, Japan}

\author{Masafumi~Kurachi}
\affiliation{Kobayashi-Maskawa Institute for the Origin of Particles and the Universe, \\
Nagoya University, Nagoya 464-8602, Japan}

\author{Toshihide~Maskawa}
\affiliation{Kobayashi-Maskawa Institute for the Origin of Particles and the Universe, \\
Nagoya University, Nagoya 464-8602, Japan}

\author{Kei-ichi~Nagai}
\affiliation{Kobayashi-Maskawa Institute for the Origin of Particles and the Universe, \\
Nagoya University, Nagoya 464-8602, Japan}

\author{Hiroshi~Ohki}
\affiliation{Kobayashi-Maskawa Institute for the Origin of Particles and the Universe, \\
Nagoya University, Nagoya 464-8602, Japan}

\author{Akihiro~Shibata}
\affiliation{Computing Research Center, High Energy Accelerator Research Organization (KEK),\\ Tsukuba 305-0801, Japan}

\author{Koichi~Yamawaki}
\affiliation{Kobayashi-Maskawa Institute for the Origin of Particles and the Universe, \\
Nagoya University, Nagoya 464-8602, Japan}

\author{Takeshi~Yamazaki}
\affiliation{Kobayashi-Maskawa Institute for the Origin of Particles and the Universe, \\
Nagoya University, Nagoya 464-8602, Japan}

\collaboration{LatKMI Collaboration}
\noaffiliation

\begin{abstract}
We investigate chiral and conformal properties of the lattice QCD
 with eight flavors ($N_f=8$) through meson spectrum
using the Highly Improved Staggered Quark (HISQ) action. 
We also compare our results with those of $N_f=12$ and $N_f=4$
which we study on the same systematics.
We find that the decay constant $F_\pi$  
of the pseudoscalar meson ``pion'' $\pi$
is  non-zero, 
with its mass $M_\pi$ consistent with zero, 
both in the chiral limit extrapolation
of the chiral perturbation theory (ChPT). 
We also measure other quantities which we find 
are in accord with the $\pi$ data results:
The $\rho$ meson mass  is consistent with non-zero in the chiral limit, 
and so is the chiral condensate, 
with its value neatly coinciding with that from the
Gell-Mann-Oakes-Renner relation  in the chiral limit.
Thus our data for the $N_f=8$ QCD are consistent with the spontaneously broken chiral symmetry. 
Remarkably enough, 
while the $N_f=8$ data  near the chiral limit 
are well described by the ChPT,
those for the relatively large fermion bare mass $m_f$ 
away from the chiral limit
actually exhibit a finite-size hyperscaling relation, 
suggesting a large anomalous dimension $\gamma_m \sim 1$. 
This implies 
that there exists a remnant of the infrared conformality,
and  suggests 
that a typical technicolor (``one-family model'')  
as modeled by the $N_f=8$ QCD 
can be a walking technicolor theory 
having an approximate scale invariance 
with large anomalous dimension
$\gamma_m \sim 1$.   
 
\end{abstract}

\maketitle

\section{Introduction}
\label{sec:intro}

The origin of mass is the most urgent issue of the particle physics today.
Although the LHC has discovered a 125 GeV boson
 roughly consistent with the Standard Model (SM) Higgs boson,
there still remain many unsolved problems with the SM, 
which would require physics beyond the SM.
One of the candidates for the theory 
beyond the SM 
towards that problem is the Walking Technicolor (WTC)~\cite{Yamawaki:1985zg} 
having a large anomalous dimension $\gamma_m \simeq 1$  
and approximate scale invariance 
due to the almost  non-running (``walking'') coupling
\cite{[{Similar works without notion of anomalous dimension and scale symmetry were done: }][]Holdom:1984sk,*Akiba:1985rr,*Appelquist:1986an},
which is based on the scale-invariant gauge dynamics 
(ladder Schwinger-Dyson equation~ \cite{Maskawa:1974vs,*Maskawa:1975hx,*Fukuda:1976zb,Miransky:1984ef}).
Actually, WTC predicts~\cite{Yamawaki:1985zg,Bando:1986bg} 
a light scalar Higgs-like composite, technidilaton,  
a pseudo Nambu-Goldstone boson of the spontaneously broken approximate scale symmetry, 
which may be identified with the 125 GeV boson~\cite{Matsuzaki:2012mk,*[][{ and references therein.}]Matsuzaki:2012xx}.

The walking behavior can  in fact be realized in the ``large $N_f$ QCD'', 
QCD with  large number of (massless) flavors  $N_f$, 
which possesses the Caswell-Banks-Zaks (CBZ)  infrared fixed point (IRFP)~\cite{Caswell:1974gg,*Banks:1981nn}, 
$\alpha_* =\alpha_*(N_c,N_f)\,(<\infty)$ of the two-loop beta function, 
for
$N_f^* (\simeq 8) <N_f < N_f^{\rm (AF)}=11N_c/2 (=16.5)$ 
in such a way that $\alpha_* \rightarrow 0$ as $N_f \rightarrow N_f^{\rm (AF)}$,
where $N_f^{\rm (AF)}$ is the maximum number to keep the asymptotic freedom.
Due to the CBZ IRFP 
there exists an approximate scale invariance $\alpha(\mu) \simeq \alpha_*$ 
in the infrared region  $0<\mu<\Lambda_{\rm QCD}$ (``infrared conformality''),
while such a scale symmetry is lost 
for the ultraviolet region $\mu>\Lambda_{\rm QCD}$ 
where the coupling runs 
as in a usual asymptotically free theory.\footnote{
 The intrinsic scale $\Lambda_{\rm QCD}$ at two-loop level 
 is defined as usual by a renormalization-group-invariant scale parameter
 $\Lambda_{\rm QCD}  =\mu\cdot  \exp \left( - \int^{\alpha(\mu)} \frac{d \alpha}{\beta(\alpha)}\right)$
 such that $\frac{d \Lambda_{\rm QCD}}{d \mu}=0$, 
 where $\beta(\alpha)\equiv \partial \alpha/\partial (\ln \mu)$ is the two-loop beta function 
 instead of the one-loop one~\cite{Appelquist:1996dq}.
 }
When  $N_f$  is near $N_f^*$  
so that $\alpha_*$ is strong enough 
to trigger the spontaneous chiral symmetry breaking (S$\chi$SB), 
the exact IRFP would actually be  washed out 
by  the dynamical generation of a quark mass  $m_D\ne 0$ 
through a continuous phase transition (``conformal phase transition''~\cite{Miransky:1996pd}), 
$m_D= 0$ ($\alpha_*<\alpha_{\rm cr}$, 
or $N_f >N_f^{\rm cr} \,(>N_f^*))$ to $m_D\ne 0$ ($\alpha_*>\alpha_{\rm cr}$,
or $(N_f^*<)\, N_f <N_f^{\rm cr}$), 
in such a way (``Miransky scaling''~\cite{Miransky:1984ef}) 
that $m_D \sim \Lambda_{\rm QCD} \cdot \exp\left(-\pi/\sqrt{\alpha_*-\alpha^{\rm cr}}\right) \ll \Lambda_{\rm QCD}$ 
for $\alpha_* \simeq \alpha_{\rm cr}$ ($N_f \simeq N_f^{\rm cr}$),  
where $\alpha_{\rm cr}$ is the critical coupling 
for the S$\chi$SB and $N_f^{\rm cr}$ the critical number of flavors 
such that $\alpha_*(N_c, N_f^{\rm cr}) =\alpha_{\rm cr}$.
The critical number $N_f^{\rm cr}$ was estimated as $N_f^{\rm cr} \simeq 4 N_c \simeq 12$~\cite{Appelquist:1996dq} 
by comparing the two-loop value of the CBZ IRFP  
with the critical coupling of  
the ladder SD equation analysis~\cite{Maskawa:1974vs}: 
$\alpha_*(N_c,N_f^{\rm cr})=\alpha_{\rm cr}\, (=\pi/(3 C_2)=\pi/4)$. 
Now, for  $N_f (<N_f^{\rm cr})$ very close to $N_f^{\rm cr}$,  
the dynamical mass $m_D \,(\ne 0)$ 
could be much smaller than the intrinsic scale 
$m_D\ll \Lambda_{\rm QCD}$, 
in sharp contrast to the usual QCD 
where $m_D ={\cal O} (\Lambda_{\rm QCD})$,  
so that the approximate conformality $\alpha(\mu) \simeq \alpha_*$ still remains 
in the wide infrared region $m_D<\mu<\Lambda_{\rm QCD}$ 
as an impact of the would-be IRFP. Such a "remnant of conformality" should appear 
in low-energy quantities. This is the case for the WTC, 
with the intrinsic scale $\Lambda_{\rm QCD}$ being identified 
with the ``ultraviolet'' cutoff $\Lambda$ of the WTC usually taken 
as the Extended Technicolor (ETC) scale $\Lambda_{\rm ETC}$,
and will be the focus of our interest 
in this paper.
  
Although the above results from the two-loop perturbation combined 
with the ladder approximation are very suggestive,  
the relevant dynamics is obviously of  non-perturbative nature,
we would need fully non-perturbative studies. 
Among others the lattice simulations developed in the lattice QCD 
would be the most powerful tool to investigate the walking behavior of the large $N_f$ QCD. 
Actually, there were some pioneering works on the large $N_f$ QCD
in somewhat different contexts~\cite{Iwasaki:1991mr,Iwasaki:2003de,Brown:1992fz,Damgaard:1997ut}, 
and more recently there have been many lattice studies 
towards the above problem\cite{[{See for example: }][]Giedt:2012it,*Neil:2012cb}.
The immediate issues are: 
What is the critical number $N_f^{\rm cr}$?
What is the signatures of the walking theory on the lattice?
In particular, the above two-loop/ladder studies would suggest 
that the walking theory 
if existed might 
be in between $N_f=8$ and $N_f=12$.
As to  $N_f=12$  
there have been many analyses 
including those of ourselves 
which are consistent with the theory 
being inside the conformal window~\cite{Iwasaki:1991mr,Iwasaki:2003de,Appelquist:2007hu,*Appelquist:2009ty,Deuzeman:2009mh,Hasenfratz:2010fi,*Hasenfratz:2011xn,DeGrand:2011cu,Ogawa:2011ki,Lin:2012iw,Aoki:2012eq,*Aoki:2012kr,*Aoki:2012yd,Itou:2012qn,Iwasaki:2012kv,Ishikawa:2013wf}, 
although some works prefer the S$\chi$SB phase~\cite{Fodor:2009wk,Jin:2010vm}. 
There were also simulations on $N_f=10$~\cite{Hayakawa:2010yn} 
consistent with the infrared conformality.   
We thus are interested in $N_f=8$ as a candidate for the walking theory. 

Actually, the $N_f=8$  is particularly interesting from the model-building point of view~\cite{[{See for a review: }]Farhi:1980xs}: 
A  typical technicolor model  is the so-called 
one-family model (Farhi-Susskind model~\cite{Farhi:1979zx})  
which has a one-family of colored 
and uncolored weak-doublets techni-fermions (techni-quarks and  techni-leptons)
corresponding to each family of the SM quarks and leptons. 
It can embed the technicolor gauge 
and the gauged three generations of the SM fermions 
into a single gauge group (ETC) 
and thus is the most straightforward way to accommodate the techni-fermions 
and the SM fermions into a simple scheme to give mass to the SM fermions.  
Thus if the $N_f=8$ turns out to be a  walking theory, 
it would be a great message for the phenomenology,
which is to be tested by the on-going LHC. 
Actually, the techni-dilaton~\cite{Yamawaki:1985zg,Bando:1986bg} 
in the WTC for the one-family model
 is consistent with the present LHC data for 125 GeV
 boson in a ladder analysis \cite{Matsuzaki:2012mk} and  in
 holographic estimate \cite{Matsuzaki:2012xx} 
 \footnote{
As to the immediate questions about the problem with the S, T parameters, see, for example, discussions in Ref.~\cite{Matsuzaki:2012gd}.
} .

\begin{figure} 
\includegraphics[scale=0.45]{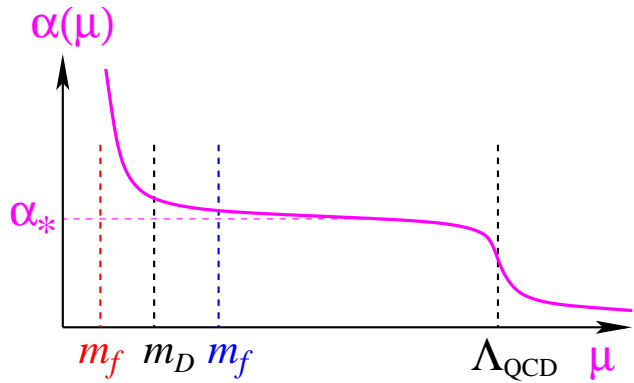}
\caption{
Schematic two-loop/ladder picture of  the gauge coupling of the
massless large $N_f$ QCD as a walking gauge theory 
in the S$\chi$SB phase 
near the conformal window. 
$m_D$ is the dynamical mass of the fermion 
generated by the S$\chi$SB.
The effects of  the bare mass of the fermion $m_f$  
would be qualitatively different depending on the cases:  
Case 1: $m_f \ll m_D$ (red dotted line) well described by ChPT, 
and Case 2: $m_f \gg m_D$ (blue dotted line) 
well described by the hyper scaling.
}
\label{fig:walking}
\end{figure}

If $N_f=8$ is a walking theory desired for the WTC,
it should be inside the S$\chi$SB phase $N_f =8 <N_f^{\rm cr}$ ($m_D\ne 0$)
and at the same time be close to the phase boundary 
with the conformal window $N_f > N_f^{\rm cr}$ ($m_D=0$) 
such that $m_D \ll \Lambda_{\rm QCD}$.
Now the lattice simulations we are making
contain several scale-symmetry breaking parameters, 
the fermion bare mass $m_f$ 
as well as  a finite box $L^3$ and lattice spacing $a$,
which do not exist in the continuum theory 
we are interested in. 
Among others the fermion bare mass $m_f$ obviously distorts 
the ideal behavior of the breaking of the scale symmetry 
in a way similar to the continuum theory.
Then, disregarding the effects of the lattice parameters $L$ and $a$ for the moment\footnote{In our simulation 
we use the parameter region
where the effect of the system size is subdominant 
compared to the mass effect.
This strategy is different from the
one which is advocated by the authors of
Refs.~\cite{Iwasaki:2012kv,Ishikawa:2013wf}.},
we may imagine  possible effects of the fermion bare mass 
on the walking coupling of our target of study
as in Fig.~\ref{fig:walking},
which is suggested by the two-loop/ladder analysis. \\
Case 1. $m_f \ll m_D\ll \Lambda_{\rm QCD}$ 
(red dotted line in Fig. \ref{fig:walking}): 
The chiral perturbation theory should hold
in a way similar to the real-life QCD with light quarks. \\  
Case 2. $m_D\ll m_f \ll \Lambda_{\rm QCD}$ 
(blue dotted line  in Fig. \ref{fig:walking}): 
The conformal hyperscaling relation should hold approximately 
with a large anomalous dimension $\gamma_m \simeq 1$. \\ 
Actually, the S$\chi$SB order parameter to be measured on the lattice
is not $m_D$ 
but would be the decay constant $F_\pi$ of the Nambu-Goldstone boson $\pi$
 extrapolated to the chiral limit:
$F=F_\pi (m_f=0)$ which would be expected 
roughly the same as $m_D$:  $ m_D={\cal O} (F)$.
%

There is a caveat about the approximate hyperscaling relation 
to be expected in the Case 2 ($m_D \ll m_f \ll \Lambda_{\rm QCD}$ ): 
There are two infrared mass parameters $m_D$ and $m_f$ 
which violate the infrared conformality 
and hence the possible hyperscaling relations 
for the physical mass quantities 
measured from the spectrum should not be universal 
but do depend on both of them in non-universal ways, 
in sharp contrast to the hyperscaling relation 
in the conformal window 
where all the mass parameters from the spectra
reflects the deformation 
by the unique infrared scale-violating parameter $m_f$  
in a universal way. 
In particular, when $m_f$ is getting close to the region in Case 1,
where $\pi$ mass $M_\pi$ and the other quantities such as
$\rho$ mass $M_\rho$ and $F_\pi$ behave qualitatively different towards
the chiral limit: $M_\pi\to 0$ while the others remain non-zero.
  
To date,
some groups
carried out lattice studies on 8-flavors,
with Wilson fermions 
\cite{Iwasaki:1991mr,Iwasaki:2003de,Iwasaki:2012kv,Ishikawa:2013wf}
and with staggered fermions
\cite{Brown:1992fz,Appelquist:2007hu,Deuzeman:2008sc,Fodor:2009wk,Jin:2010vm,Miura:2012zqa,Petropoulos:2012mg,Cheng:2013eu}.
The Refs.~\cite{Iwasaki:1991mr,Iwasaki:2003de,Iwasaki:2012kv,Ishikawa:2013wf} concluded the $N_f=8$ is
in the conformal window, but 
Refs.
\cite{Brown:1992fz,Appelquist:2007hu,Deuzeman:2008sc,Fodor:2009wk,Jin:2010vm,Miura:2012zqa}
concluded
that the $N_f=8$ resides on the chiral broken phase.
Even if $N_f=8$ is in the chiral broken phase,
it has not been investigated whether the behavior of this system is QCD like
or the walking with the large anomalous mass dimension.

In this paper 
we  study the meson spectrum by simulating the $N_f=8$ QCD, 
based on yet another lattice fermion, 
Highly Improved Staggered Quark (HISQ)~\cite{Follana:2006rc,*Bazavov:2010ru}, 
applied to $N_f=8$ for the first time. 
Preliminary reports were given in Ref.~\cite{Aoki:2012ep,*Aoki:2013dz}.
HISQ action improves the behavior towards the continuum limit through
the improvement of the flavor symmetry.
The salient feature of our collaboration  is 
that we have been investigating $N_f= 4, 8, 12, 16$ 
on the setting of HISQ action with the same systematics
in order to study the $N_f$-dependence of the physics systematically
\cite{Aoki:2012kr,Aoki:2012eq,Aoki:2012ep,Aoki:2013dz}. 
Thus our analyses for $N_f=8$ are made in comparison 
with those for other flavors of our group.

We first show the data of the meson spectrum, $M_\pi$ and $F_\pi$, 
as well as $M_\rho$ and the chiral condensate  $\langle \bar \psi \psi \rangle$ 
for $\beta(\equiv 6/g^2)=3.8$ 
on the $L^3\times T$ lattice 
with and $L=12-36$ and $T=16-48$,
and $m_f=0.015-0.16$. 
We find the two regions of $m_f$ having qualitatively different properties:
$m_f=0.015-0.04$ and $m_f=0.05-0.16$.
We analyze the data based on the Chiral Perturbation Theory (ChPT)~\cite{Gasser:1983yg,*Gasser:1984gg}, 
for small $m_f$:  $m_f=0.015-0.04$ 
(roughly corresponding to Case 1 in Fig. \ref{fig:walking} in the above). 
We find 
that the ChPT analysis is self-consistent 
and find a result consistent with non-zero value of $F$ and $M_\rho$  
and vanishing of $M_\pi$ in the chiral limit extrapolation 
based on the ChPT 
(we also estimate the effects of the chiral logarithm).
The chiral condensate is also non-zero value  
in the chiral limit extrapolation, 
which neatly coincides with the Gell-Mann-Oakes-Renner (GMOR) relation 
obtained from the $\pi$ data in the chiral limit extrapolation.

As to the large $m_f$ ($m_f=0.05-0.16$)
(roughly corresponding to the Case 2 in Fig.\ref{fig:walking}),
we find the finite-size hyperscaling (FSHS)
\cite{DeGrand:2009hu,DelDebbio:2010ze,DelDebbio:2010hu,DelDebbio:2010hx}
holds in this region,
when we take into account mass corrections to the FSHS.
Note that such corrections were
sizable \cite{Aoki:2012eq} in the large mass region
even for $N_f=12$, which
are consistent to be in the
conformal window.
From the hyperscaling analysis for such a large $m_f$,
we find a large anomalous mass-dimension $\gamma_m \sim 1$ 
consistent with that desired by the WTC.
This implies 
that there exists a remnant of the infrared conformality
where the spontaneous chiral symmetry breaking (S$\chi$SB) effects 
are negligible 
compared with the mass deformation $m_f$.
It is the first time that
the hyperscaling relation is observed  in a theory with S$\chi$SB.

The S$\chi$SB feature of $N_f=8$ data near the chiral limit 
are found to be qualitatively similar to those of the $N_f=4$ case:
We actually find $N_f=4$ data indicate robust signals of S$\chi$SB phase. 
On the other hand, our $N_f=4$ data indicate no trace of the hyperscaling relation  
for large $m_f$ region in sharp contrast to $N_f=8$ data. 
The $N_f=8$ result is also contrasted with the $N_f=12$ 
where our previous study concluded that the ChPT analysis with our data
was not self-consistent, while the FSHS relation held consistently with
the infrared conformality. 

This suggests that a typical technicolor (``one-family model'')  
as modeled by the $N_f=8$ QCD can be a walking technicolor theory 
having an approximate scale invariance with large anomalous dimension.

This article is organized as follows:
Sec.~\ref{sec:sim} presents 
our lattice simulation setup, calculation of observables,
analysis method, and the results of the crude analysis of our data.
Sec.~\ref{sec:chpt} shows the analysis based on the ChPT to show that
$N_f=8$ is actually in the S$\chi$SB phase.
Sec.~\ref{sec:FSHS} is to study the remnants of conformality.
Sec.~\ref{sec:summary} is devoted to the summary and discussion.
Appendices \ref{app:data} and \ref{app:nf4} summarize 
detailed numerical results for $N_f=8$ and 4, respectively.
In Appendix \ref{app:log} we estimate chiral log corrections
in $N_f=8$.
We analyze FSHS in an alternative method   in Appendix \ref{app:pgamma} .

\section{Lattice simulation and the results}
\label{sec:sim}
\subsection{Lattice setup}
\label{subsec:setup}

In our simulation,
we use the tree-level Symanzik  gauge action
and the highly improved staggered quark (HISQ) action~\cite{Follana:2006rc}
without the tadpole improvement 
and the mass correction in the Naik term\cite{Bazavov:2011nk}.
It is expected that
the flavor symmetry in the staggered fermion 
and the behavior towards the continuum limit are improved
by HISQ improvement.
We carry out the simulation
by using the standard Hybrid Monte-Carlo (HMC) algorithm
using MILC code version 7~\cite{MilcCode}
with some modifications to suit our needs.
One of the modifications is
the Hasenbush mass preconditioning~\cite{Hasenbusch:2001ne}
to reduce the large computational cost of
the configuration generation at the smaller $m_f$.
We measure the mass of the pion $M_\pi$,
$\rho$-meson $M_\rho$
and the decay constant of the pion $F_\pi$	
and the chiral condensate $\langle \bar{\psi}\psi \rangle$
as the basic observable
to explore the large-$N_f$ QCD.

The simulation in the preliminary report~\cite{Aoki:2012ep}, which includes the study of the 
anomalous dimension, for $N_f=8$ is carried out 
at $\beta(=6/g^2)$=3.6, 3.7, 3.8, 3.9, 4.0
for various quark masses and on various lattices, $L^3 \times T$,
where $L$ is the spatial size and $T$ the temporal size.
We need to choose as small value of $\beta$ as possible to obtain a large enough 
physical volume to minimize the finite-volume effect.
From the global survey mentioned above, we found that $\beta < 3.8$ 
is too strong to carry out  the HMC simulation with HISQ. 
Therefore, we choose $\beta=3.8$ in this article.

Note that the aspect ratio is kept fixed as $T/L = 4/3$,
in which $L=12$, 18, 24, 30 and 36.
The boundary condition in the spatial direction is the periodic
and the one in the temporal direction is the anti-periodic for fermions.
We take more than 700 trajectories for the ensemble
with 4-5 steps for saving the configuration.
The error analysis is performed 
with the standard jackknife analysis 
having the suitable bin size, 40 trajectories.
In the following analyses the error of the fit result is
estimated from the standard deviation of least squares coefficients.
See the details of the simulation parameter 
in Tables~\ref{tab:1}--\ref{tab:5}.

We also generate
gauge configurations 
for $N_f=4$.
From $\beta=$3.6, 3.7, 3.8 which were investigated in the preliminary study~\cite{Aoki:2012ep},
here we focus on $\beta=3.7$, which is appropriate
for our purpose,
with high accuracy
on $12^3 \times 18$, $16^3 \times 24$ and $20^3 \times 30$.
See Appendix~\ref{app:nf4} for details.

\subsection{Calculation of observables}
\label{subsubsec:obs}

We measure the two-point correlation functions 
of the staggered bilinear pseudoscalar operator 
which corresponds to the Nambu-Goldstone (NG) mode 
associated with the chiral symmetry of the staggered fermions.
The corresponding spin-flavor structure is $(\gamma_5\otimes \xi_5)$,
denoted by ``PS'' in Ref.~\cite{Bowler:1986fw}.
The random wall source is used for the quark operator for the bilinear,
which becomes a noisy estimator of the point bilinear operator with
spatial sum at a given time slice $t_0$.
We combine quark propagators 
solved with periodic and antiperiodic boundary conditions 
in the temporal direction~(see, e.g., Ref.~\cite{Blum:2001xb}),
which is denoted by ``P+AP'' in this article.
In this well-known technique,  
the temporal size is effectively doubled, 
which enables us to have sufficient range for the fitting.
Denoting such a $\pi$ correlator as $C_{\rm PS}(t)$, 
this behaves as the following expression in the staggered fermion
with P+AP prescription:
\begin{equation}
C_{\rm PS}(t)=C \left(  e^{-M_{\pi} t}  + e^{-M_{\pi}(2T- t)} \right) + B (-1)^t  \, ,
\label{eq:cosh} 
\end{equation}
where $B$ is the constant term in the oscillation mode
and $M_\pi$ is the mass of NG-pion mode,
and $C$ is the amplitude relating to the decay.
Here we use
\begin{equation}
\tilde{C}_{\rm PS}(2t)=C_{\rm PS}(2t)/2 +C_{\rm PS}(2t-1)/4+C_{\rm PS}(2t+1)/4 .
\label{eq:tildeps}
\end{equation}
This linear combination kills the constant oscillation mode, 
which could originate from the single quark line wrapping 
around the antiperiodic temporal boundary. 
The mass of NG-pion is obtained by the fit of the two-point correlators 
of $\widetilde{C}_{\rm PS}$ 
from a random source with double period 
by a fit function with the fit range $[t_{min}, T]$,
\begin{equation}  \label{eq:PS}
\tilde{C}_{\rm PS}(2t)={\tilde C} \left (e^{-M_{\pi}2t}  + e^{-M_{\pi}(2T-2t)} \right),
\end{equation}
where $2 {\tilde C} = C \left(1+ \cosh(M_\pi) \right)$. 

The pseudoscalar decay constant, $F_\pi$,  is obtained 
through the matrix element of the pseudoscalar operator,
\begin{equation}
 F_\pi= \frac{ m_f}{M_\pi^2} \langle  0| P^a(0) |\pi^a ;\vec{p} \rangle \,,
 \label{eq:fpidef}
\end{equation}
by using partially conserved axial current (PCAC) 
relation\footnote{We use the convention as $F_\pi=\sqrt{2}f_\pi$, 
where $f_\pi=93$[MeV] in the real-life QCD.}.

We measure $M_\rho$ from the staggered vector meson operator
$(\gamma_i\gamma_4\otimes\xi_i\xi_4)$, denoted by PV in
Ref.~\cite{Bowler:1986fw}. 
The asymptotic form of the PV correlator at large $t$ may be written as 
\begin{equation} \label{eq:rho}
C_{\rm PV}(t)=C_1( e^{-M_\rho t} +e^{-M_\rho(2T-t)}) 
+C_2 (-1)^t ( e^{-M_{a_1} t} +e^{-M_{a_1}(2T-t)}) 
\end{equation}
where $M_{a_1}$ corresponds to the mass of the axialvector meson
which is the parity partner mode of PV mode in the staggered fermion.
Since there exists a constant mode
due to the wrapping-around effect, 
we use  
\begin{equation}  
\tilde{C}_{\rm PV}(2t)=C_{\rm PV}(2t)/2 +C_{\rm PV}(2t-1)/4+C_{\rm PV}(2t+1)/4 \,.
\label{eq:tildepv}
\end{equation}  
Therefore, 
\begin{equation}  \label{eq:PV}
\tilde{C}_{\rm PV}(2t)=\tilde{C_1} (e^{-M_{\rho}2t}  + e^{-M_{\rho}(2T-2t)})
+\tilde{C_2}( e^{-M_{a_1}2 t} +e^{-M_{a_1}(2T-2t)}) ,
\end{equation}
where $2 \tilde{C_1}=C_1 \left(1 + \cosh(M_\rho) \right) $
and $2 \tilde{C_2}=C_2 \left(1 - \cosh(M_{a_1}) \right) $. 
Even in the case of $M_\rho \simeq  M_{a_1}$ and $C_1 \simeq C_2$,
we have $\tilde{C_1} \gg \tilde{C_2}$
for our typical value of $M_\rho$.
Then Eq.~(\ref{eq:PV}) can be approximated to 
the simple cosh function
of the two-point correlators of $\tilde{C}_{\rm PV}$: 
\begin{equation}  \label{eq:PVcosh}
\tilde{C}_{\rm PV}(2t)=\tilde{C_1} (e^{-M_{\rho}2t}  + e^{-M_{\rho}(2T-2t)}) \,,
\end{equation}
and we obtain $M_\rho$.

Besides these main channels,  
we study the masses of mesons interpolated from local operators,
a non-NG channel $(\gamma_5\gamma_4\otimes\xi_5\xi_4)$ denoted by ``SC'', 
and a vector meson $(\gamma_i\otimes\xi_i)$ denoted by ``VT'' in Ref.~\cite{Bowler:1986fw},
by which we will show that  the flavor-symmetry breaking is small in HISQ.
These masses are obtained from corner wall source correlator.

\begin{figure}[!h] 
\makebox[.45\textwidth][r]{\includegraphics[scale=.35]{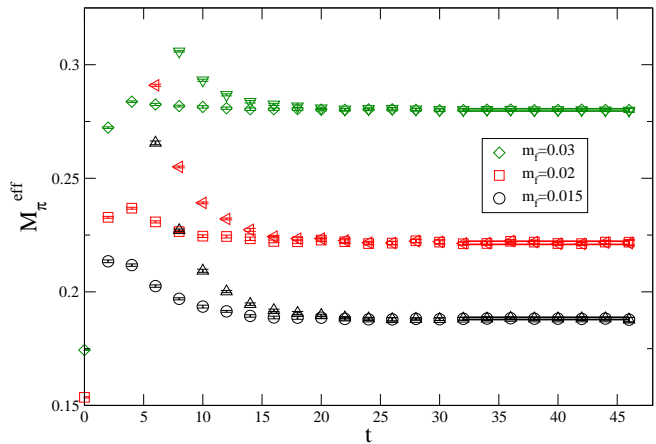}}
\caption{
Effective masses of PS meson, $M_\pi^{\rm eff}$,  at L=36.
Triangles and other symbols denote results from
point sink correlators with
random wall source and 
corner wall source, respectively.
Fit results with error band obtained from random wall source correlator 
are also plotted by solid lines.
}
\label{fig:meff}
\end{figure}

The effective masses 
are calculated through 
$\tilde{C}_{\rm PS}(2t)$ defined in
Eq.~(\ref{eq:tildeps}). 
Fig.~\ref{fig:meff}  shows  typical examples of PS channel for the largest volume. 
The horizontal lines show the results of fitting
$\tilde{C}_{\rm PS}(2t)$ with Eq.~(\ref{eq:PS}) with $32\le 2t \le 48$, 
where plateau is observed.
%
All the fit results are summarized in Tables~\ref{tab:1}--\ref{tab:5}.

We also calculate the chiral condensate $\langle {\bar \psi} \psi \rangle$
normalized for a single Dirac flavor which can be obtained 
through 
\begin{equation}
\langle {\bar \psi}(x) \psi(x) \rangle = \frac{1}{4} {\rm Tr} \left[ D_{HISQ}^{-1}(x,x) \right],
 \label{eq:pbp}
\end{equation}
where $ D_{HISQ}(x,y) $ is the single species (four flavor) staggered
Dirac operator for HISQ.
Here an average over the space-time $x$ is calculated through a stochastic method.

\subsection{Analysis methods}
\label{subsubsec:analysis}

We performed the analysis 
based on  the chiral perturbation  theory (ChPT) 
and the (finite-size) hyperscaling,
as explained in the following;
If the system is in the spontaneous chiral symmetry broken (S$\chi$SB) phase,
physical quantities in the spectroscopy, $M_H$ for $H=\pi, \rho, \cdots$ and $F_\pi$,  
are described by the ChPT.
The mass and decay constant of $\pi$ depend on $m_f$ 
up to chiral log as
\begin{equation} 
M_\pi^2  = C_1^\pi  m_f + C_2^\pi m_f^2 + \cdots \, ,\quad F_\pi = F + C_1^F m_f + C_2^F m_f^2 + \cdots \, ,
\label{eq:chpt}
\end{equation}
where $F$ is the value in the chiral limit.

On the other hand,
if the theory is in the conformal window,
$M_H$ and $F_\pi$ obey the conformal hyperscaling\cite{Miransky:1998dh}
\begin{equation}
M_H \propto m_f^{\frac{1}{1+\gamma_\ast}} \,, \quad F_\pi \propto m_f^{\frac{1}{1+\gamma_\ast}} \, ,
\label{eq:hs}
\end{equation}
where $\gamma_\ast$ denotes the mass anomalous dimension $\gamma_m$
at the infrared fixed point
and its value is universal for all channels.
On the finite volume
$M_H$ and $F_\pi$ are described by the finite size hyperscaling (FSHS) \cite{DeGrand:2009hu,DelDebbio:2010ze,DelDebbio:2010hx,DelDebbio:2010hu}
on dimension-less quantities
\begin{equation}
 \xi_p\equiv L M_p\ \ \ \ \mbox{for $p=\pi$ or $\rho$},
\label{eq:xip}
\end{equation}
or
\begin{equation}
 \xi_F\equiv L F_\pi,
\label{eq:xiF}
\end{equation}
given as
\begin{equation}
\xi_H = {\cal F}_H(L m_f^{\frac{1}{1+\gamma_\ast}}) \, ,
\label{eq:FsHs}
\end{equation}
where $H=\pi$, $\rho$ or $F$.
The function, ${\cal F}_H$, is a some function (unknown {\it a priori})
of the scaling variable $X=L m_f^{\frac{1}{1+\gamma_\ast}}$.

\subsection{Results}
\label{subsec:spect}

\begin{figure}[!h]
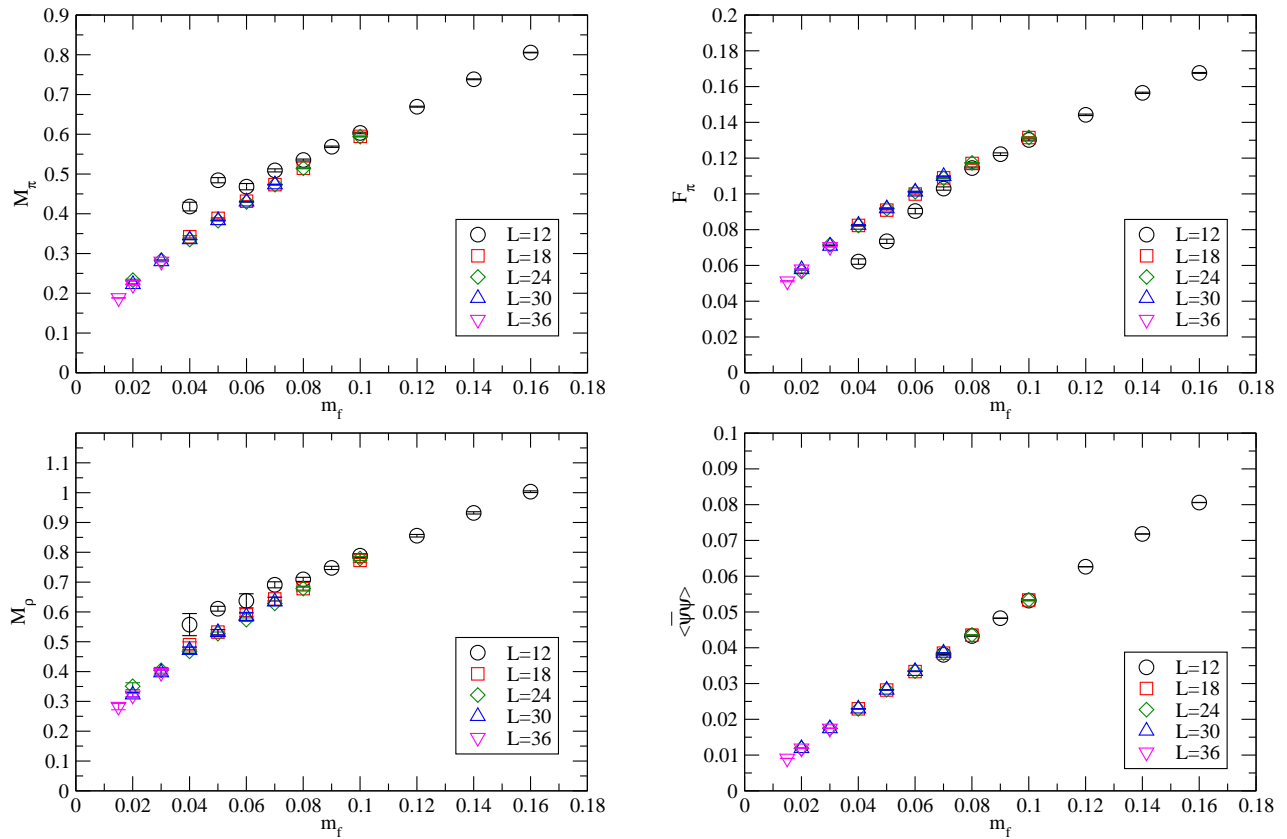
 
\makebox[.49\textwidth][r]{\includegraphics[scale=.45]{nf8_figures/fig-mpi-mf.eps}}
\makebox[.49\textwidth][r]{\includegraphics[scale=.45]{nf8_figures/fig-fpi-mf.eps}}
\makebox[.49\textwidth][r]{\includegraphics[scale=.45]{nf8_figures/fig-mpv-mf.eps}}
\makebox[.49\textwidth][r]{\includegraphics[scale=.45]{nf8_figures/fig-pbp-mf.eps}}
\caption{
Raw data of observables as a function of $m_f$ for 
$M_\pi$ (top left), $F_\pi$ (top right), $M_\rho$ (bottom left) and
$\langle {\bar \psi} \psi \rangle$ (bottom right).
}
\label{fig:nf8}
\end{figure}

Spectral quantities, 
such as 
$M_\pi$, $M_\rho$, $F_\pi$, $\langle {\bar \psi} \psi \rangle$,
are calculated on the gauge field ensembles 
for the $N_f=8$ QCD at $\beta=3.8$,
as described in Sect.~\ref{sec:sim}.
The $m_f$ dependence of the results is shown in Fig.~\ref{fig:nf8}.
Large finite size effect is observed for smaller $m_f$ region on $L=12$. 

%

\begin{figure}[!h] 
\makebox[.49\textwidth][r]{\includegraphics[scale=.40]{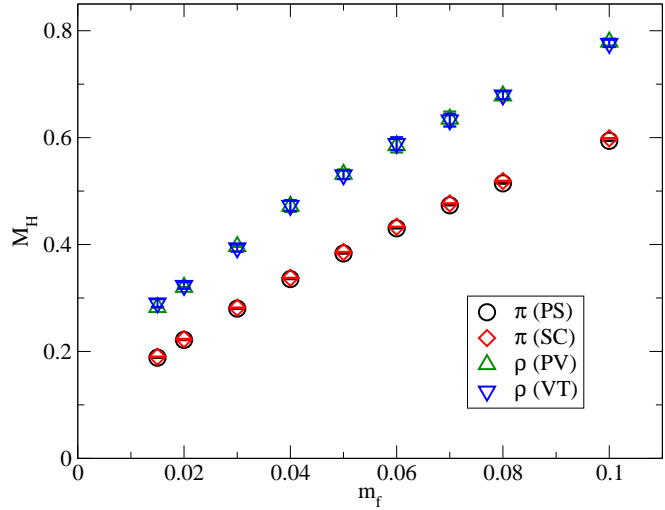}}
\caption{
Comparisons of $M_\pi$ and $M_{SC}$, and of $M_{\rho (PV)}$ and $M_{\rho (VT)}$
as a function of $m_f$
with largest volume data at each $m_f$.
}
\label{fig:flbr}
\end{figure}

The expected good flavor symmetry in HISQ action 
is actually observed in near degeneracy of
PS and SC, and of PV and VT.
See Fig.~\ref{fig:flbr}.
%

Before giving the in-depth analyses in the following sections,
let us perform some crude analysis here.
Spontaneous chiral symmetry breaking leads to non-zero $F_\pi$ and $M_\rho$
while vanishing $M_\pi$ 
in the chiral limit. 
Thus the ratios $F_\pi/M_\pi$ and $M_\rho/M_\pi$
should diverge in the chiral limit.
On the other hand,
 in the conformal phase
the ratios should take a constant value
near the chiral limit
as implied by the hyperscaling relation in Eq.~(\ref{eq:hs}).

\begin{figure}[!h]
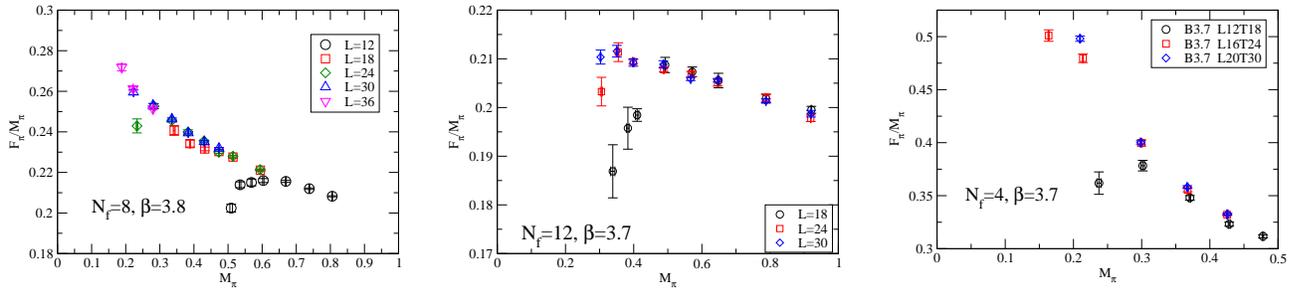
 
\makebox[.32\textwidth][r]{\includegraphics[scale=.3]{nf8_figures/fig-fpi_ov_mpi-mpi.eps}}
\makebox[.32\textwidth][r]{\includegraphics[scale=.3]{nf12_figures/fpi-pi_ratio.eps}}
\makebox[.32\textwidth][r]{\includegraphics[scale=.3]{nf4_figures/fpi-ov-mpi_Nf04_B3.7.eps}}
\caption{
$F_\pi/M_\pi$ as a function of $M_\pi$ for $N_f=8$ (left),
$N_f=12$ at $\beta=3.7$ (center) in Ref.~\cite{Aoki:2012eq},
and $N_f=4$ at $\beta=3.7$ (right).
}
\label{fig:fpimpi}
\end{figure}

\begin{figure}[!h]
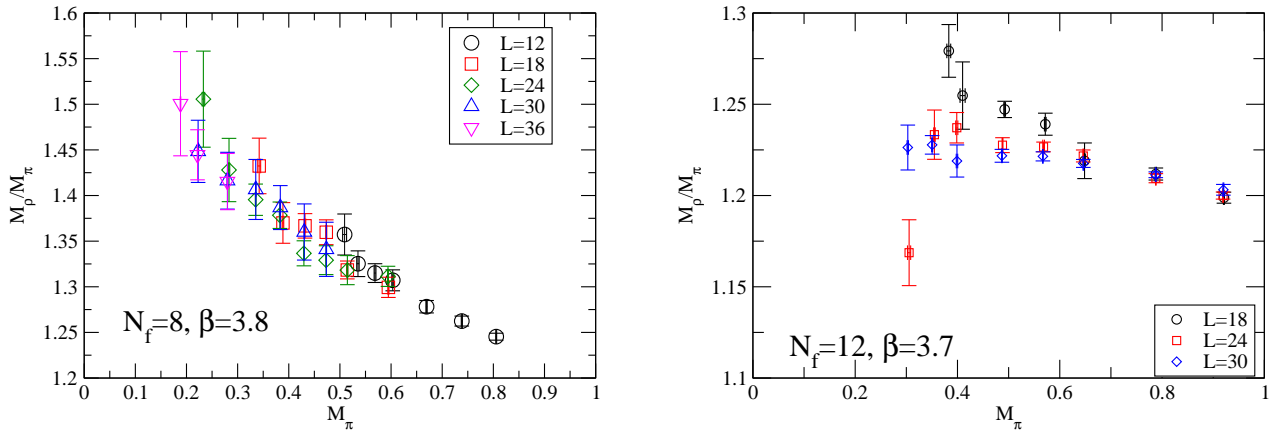
 
\makebox[.49\textwidth][r]{\includegraphics[scale=.45]{nf8_figures/fig-mpi_ov_mpv-mpi.eps}}
\makebox[.49\textwidth][r]{\includegraphics[scale=.45]{nf12_figures/rho-pi_ratio.eps}}
\caption{
$M_\rho/M_\pi$ as a function of $M_\pi$ for $N_f=8$ (left)
and $N_f=12$ at $\beta=3.7$ (right) in Ref.~\cite{Aoki:2012eq}.
}
\label{fig:mrhompi}
\end{figure}

Now look at Figs.~\ref{fig:fpimpi} and ~\ref{fig:mrhompi} 
which show 
that the ratio increases monotonically
towards the chiral limit,
if one takes the largest volume data at each $M_\pi$.
This resembles the $N_f=4$ case 
where S$\chi$SB is clearly observed
and shows clear contrast against the same plot for $N_f=12$
which are consistent with conformality.
This strongly suggests that
$N_f=8$ QCD is in S$\chi$SB phase.
In order to further study  the chiral property of $N_f=8$,
we carry out ChPT analysis 
in the next section.
%
%
%

\section{Chiral perturbation Theory analysis}
\label{sec:chpt}

\begin{figure}[!h]
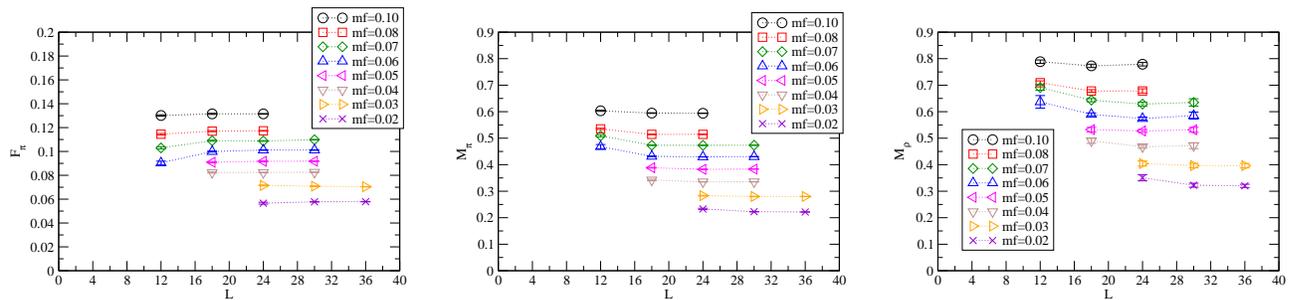
 
\makebox[.32\textwidth][r]{\includegraphics[scale=.3]{nf8_figures/fig-fpi-L.eps}}
\makebox[.32\textwidth][r]{\includegraphics[scale=.3]{nf8_figures/fig-mpi-L.eps}}
\makebox[.32\textwidth][r]{\includegraphics[scale=.3]{nf8_figures/fig-mpv-L.eps}}
\caption{
$F_\pi$ (left),  $M_\pi$ (center) and $M_\rho$
 (right) as functions of $L$.
}
\label{fig:finitevol}
\end{figure}

In order to carry out the ChPT analysis,
the finite volume effect has to be taken into account.
Fig.~\ref{fig:finitevol} shows 
the spatial size $L$ dependence of $F_\pi$, $M_\pi$ and $M_\rho$
plotted from the data on Tables~\ref{tab:1}, ~\ref{tab:2},  ~\ref{tab:3}, ~\ref{tab:4} and ~\ref{tab:5}.
We find that 
the data on the largest two volumes,
at least in this $m_f$ range,
are consistent with each other.
For the lightest $m_f$,
since there is only one volume data,
we cannot study the finite size effect.
We, however, find that
the $L M_\pi$ in the lightest $m_f$ 
is bigger than the one of $m_f=0.02$
at $L=30$ 
(see, Table~\ref{tab:4}) 
where the finite size effect is negligible.
In the following analysis
we understand that there is no finite size effect
in the lightest $m_f$.
Therefore, 
we use the data on the largest lattice
at each $m_f$ 
and perform the infinite volume ChPT analysis.
%

\subsection{Quadratic fit  of  $F_\pi$}
\label{subsec:fpiquad}

\begin{figure}[!h] 
\makebox[.49\textwidth][r]{\includegraphics[scale=.5]{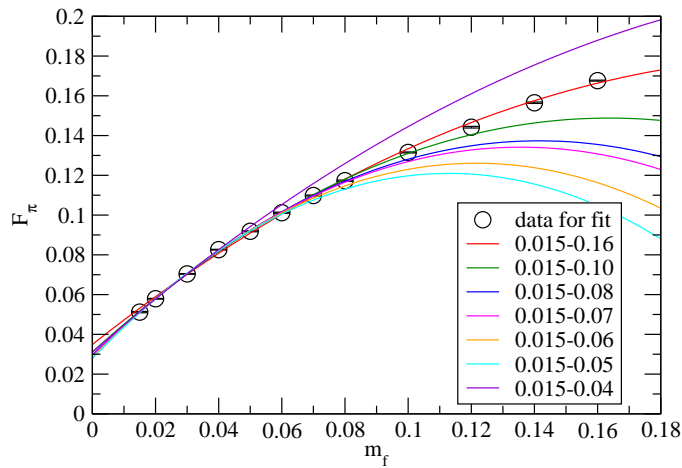}}
\caption{
Results of quadratic fit of $F_\pi$ for various fit ranges.
}
\label{fig:nf8fpichpt}
\end{figure}

Let us analyze the behavior of $F_\pi$,
towards the chiral limit.
Fig.~\ref{fig:nf8fpichpt} shows the result of $F_\pi$
at each $m_f$. 
We perform the quadratic fit for $F_\pi$
by varying the fit range of $m_f$.
(We will estimate the effect of the chiral log corrections later.)
%
The quadratic fit result of $F_\pi$ is written on the Table~\ref{tab:quadfit}.
\begin{table}
\caption{Results of chial fit of $F_\pi$ with $F_\pi=F+C_1 m_f + C_2
 m_f^2$ for various fit ranges.
}
\label{tab:quadfit}
\begin{ruledtabular}
\begin{tabular}{cddddc}
\multicolumn{1}{c}{fit range ($m_f$)} &
\multicolumn{1}{c}{$F$} &
\multicolumn{1}{c}{$\mathcal{X}(m_f^{\min}=0.015)$} &
\multicolumn{1}{c}{$\mathcal{X}(m_f=m_{\max})$} &
\multicolumn{1}{c}{$\chi^2/{\rm dof}$} &
\multicolumn{1}{c}{dof} \\
\hline
0.015--0.04   &  0.0310(13)&  3.74  &  11.80 &  0.46 &1\\
0.015--0.05   &  0.0278(8)  &  4.64 &  19.28 &  5.56  & 2\\
0.015--0.06   &  0.0284(6)  &  4.44 &  23.2 &  4.09   & 3\\
0.015--0.07   &  0.0293(5)  &  4.18 &  26.5 &  4.46  & 4\\
0.015--0.08   &  0.0296(4)  &  4.10 &  30.6 &  4.06 & 5\\
0.015--0.10   &  0.0311(3)  &  3.70 &  37.0 &  7.85 & 6\\
0.015--0.16   &  0.0349((2) &  2.94 &  54.0 &  34.2 & 9\\
\end{tabular}
\end{ruledtabular}
\end{table}
As seen in Fig.~\ref{fig:nf8fpichpt} and Table~\ref{tab:quadfit},
$F$ is non-zero ($ \sim 0.03$).
Particularly for the small region,   $0.015 \leq m_f \leq 0.04$,
the polynomial fit gives the good $\chi^2$/dof ($=0.46$).
When we include the data at $m_f=0.05$,
$\chi^2$/dof jumps up.
Although this jump might be caused by the instability due to small dof=2,
the large $\chi^2$/dof persists for the range with larger masses, thus,
with the value of $\chi^2$/dof being more reliable.
This suggests that there is a bound, 
beyond which the ChPT does not describe the data well, 
and that bound is around $m_f\lesssim 0.05$.
With this consideration and the good chiral behavior observed 
for other quantities for $m_f=0.015-0.04$,
which we will see below,
we chose $m_f=0.015-0.04$
for the fitting range of all quantities.

For the consistency of the ChPT 
particularly for the large $N_f$ QCD, 
the expansion parameter~\cite{Soldate:1989fh,*Chivukula:1992gi,*Harada:2003jx}   
for the given $M_\pi$ is defined as
\begin{equation}
\mathcal{X} =N_f \left( \frac{M_\pi}{4 \pi F/\sqrt{2}} \right)^2 ,
\label{eq:X}
\end{equation}
and this quantity is required to satisfy the condition $\mathcal{X}  < 1$,
which, however,
could become easily violated 
when the simulation is made for heavy $M_\pi$ compared to $F$.
We have $\mathcal{X}  = O(1)$
in our smallest $m_f$.
Thus the ChPT is barely self-consistent 
in contrast to the case of $N_f=12$
where  $\mathcal{X}  \simeq 40$ \cite{Aoki:2012eq}.

The above analysis suggests that 
our result in $N_f=8$  is consistent with S$\chi$SB phase
with
\begin{equation}
F=0.0310(13) 
\label{eq:chiralF}
\end{equation}
up to chiral log. 
Effects of the chiral log will be discussed later.

\subsection{Quadratic fits of $M_\rho$ and $M_\pi^2$}
\label{subsec:mrhoquad}

Here,
we attempt the quadratic fit of $M_\rho$ and $M_\pi^2$
to see whether $M_\rho \neq 0$ 
and $M_\pi^2 =0$ in the chiral limit.

\begin{figure}[!h] 
\makebox[.49\textwidth][r]{\includegraphics[scale=.45]{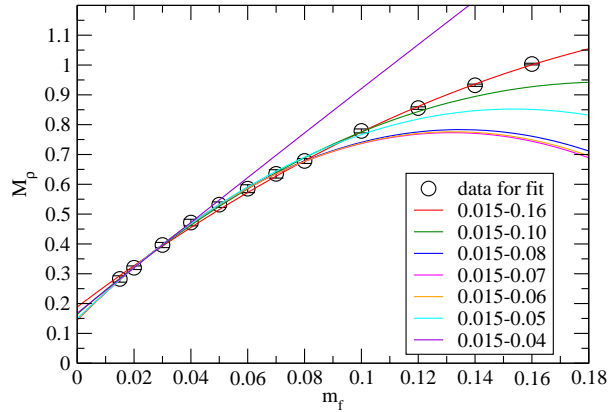}}
\caption{Results of quadratic fit of $M_\rho$ for various fit ranges.}
\label{fig:mrhoquad}
\end{figure}

Fig.~\ref{fig:mrhoquad} and Table~\ref{tab:mrhoquad}
are the quadratic fit result of $M_\rho$.
\begin{table}
\caption{Chiral fit of $M_\rho$ with 
$M_\rho=C_0^\rho+C_1^\rho m_f + C_2^\rho m_f^2$  for various fit ranges.
}
\label{tab:mrhoquad}
\begin{minipage}{.49\textwidth}
\begin{ruledtabular}
\begin{tabular}{cddc}
\multicolumn{1}{c}{fit range ($m_f$)} &
\multicolumn{1}{c}{$C_0^\rho$} &
\multicolumn{1}{c}{$\chi^2/{\rm dof}$} &
\multicolumn{1}{c}{dof} \\
\hline
 0.015--0.04  &  0.168(32)  &  0.0017 &1\\
 0.015--0.05  &  0.149(33)  &  0.098 &2 \\
 0.015--0.06  &  0.145(25)  &  0.084 &3\\
 0.015--0.07  &  0.144(20)  &  0.063 &4 \\
 0.015--0.08  &  0.146(16)  &  0.052 &5 \\
 0.015--0.10  &  0.164(12)  &  0.57 &6 \\
 0.015--0.16  &  0.189(7)   &  1.48 &9\\
\end{tabular}
\end{ruledtabular}
\end{minipage}
\end{table}
The chiral limit value of $M_\rho$ ($=C_0^\rho$) is estimated using the
fitting range $0.015 \leq m_f \leq 0.04$,
\begin{equation}
 M_\rho = 0.168(32).
  \label{eq:M_rho}
\end{equation}
%

\begin{figure}[!h]
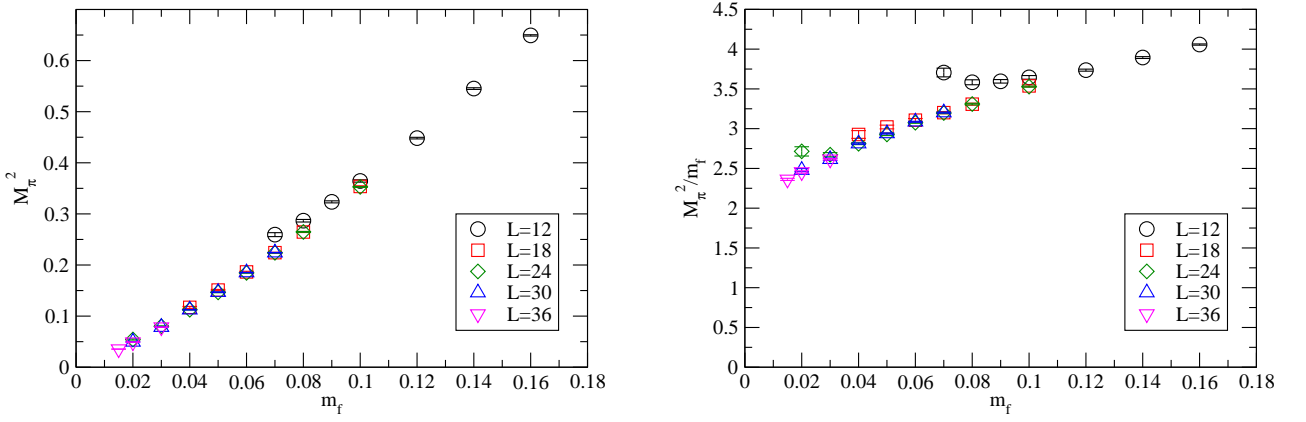
 
\makebox[.49\textwidth][r]{\includegraphics[scale=.45]{nf8_figures/fig-mpi2-mf.eps}}
\makebox[.49\textwidth][r]{\includegraphics[scale=.45]{nf8_figures/fig-mpi2_ov_mf-mf.eps}}
\caption{
$M_\pi^2$ (left) and $M_\pi^2/m_f$ (right) as functions of $m_f$.
}
\label{fig:mpi2}
\end{figure}

\begin{figure}[!h] 
\makebox[.49\textwidth][r]{\includegraphics[scale=.45]{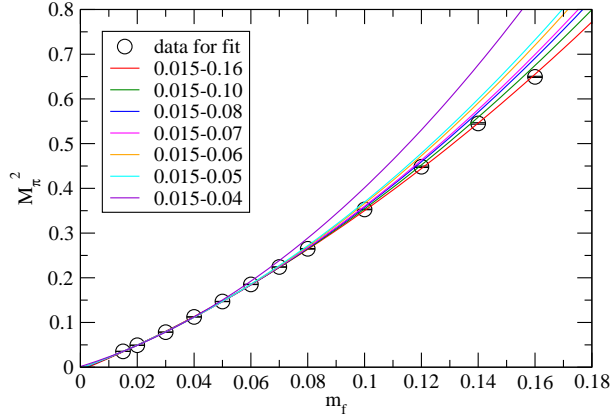}}
\caption{
Quadratic fit of $M_\pi^2$ for various fit ranges.
}
\label{fig:mpi2quad}
\end{figure}

The left panel on Fig.~\ref{fig:mpi2} shows $M_\pi^2$ 
and the right panel $M_\pi^2/m_f$ 
as a function of $m_f$.
The $M_\pi^2/m_f$ goes to constant towards the chiral limit,
which is consistent with the leading ChPT behavior.
However, the visible slope is observed,
indicating that there are higher order corrections.
This is in contrast to $N_f=4$ shown in Fig.~\ref{fig:nf4all}.
We analyze $M_\pi^2$ by  the quadratic fit with the constant term
to see whether this constant term becomes zero or not.
The result is shown in Fig.~\ref{fig:mpi2quad}.
In the fitting region $0.015 \leq m_f \leq 0.04$ 
the constant term is consistent with zero
as presented in Table.~\ref{tab:mpi2quad}.
%
%
%
\begin{table}
\caption{Chiral fit results for $M_\pi^2$ with $M_\pi^2=C_0^\pi+C_1^\pi
 m_f + C_2^\pi m_f^2$ for various fit ranges.
}
\label{tab:mpi2quad}
\begin{minipage}{.49\textwidth}
\begin{ruledtabular}
\begin{tabular}{cddc}
\multicolumn{1}{c}{fit range ($m_f$)} &
\multicolumn{1}{c}{$C_0^\pi$} &
\multicolumn{1}{c}{$\chi^2/{\rm dof}$} &
\multicolumn{1}{c}{dof} \\
\hline
 0.015--0.04  &    0.0016(13) &  1.21 &1\\
 0.015--0.05  &   -0.0017(9)  &  5.90 &2\\
 0.015--0.06  &   -0.0022(6)  &  4.18 &3\\
 0.015--0.07  &   -0.0032(5)  &  5.00 &4 \\
 0.015--0.08  &   -0.0037(5)  &  5.44&5 \\
 0.015--0.10  &   -0.0049(4)  &  7.28 &6 \\
 0.015--0.16  &   -0.0071(3)  & 14.8 &9\\
\end{tabular}
\end{ruledtabular}
\end{minipage}
\end{table}
Therefore
chiral property of $M_\rho$ and $M_\pi$ is also consistent with that of S$\chi$SB.

\subsection{Chiral condensate}
\label{subsec:cond}

\begin{figure}[!h] 
\makebox[.49\textwidth][r]{\includegraphics[scale=.45]{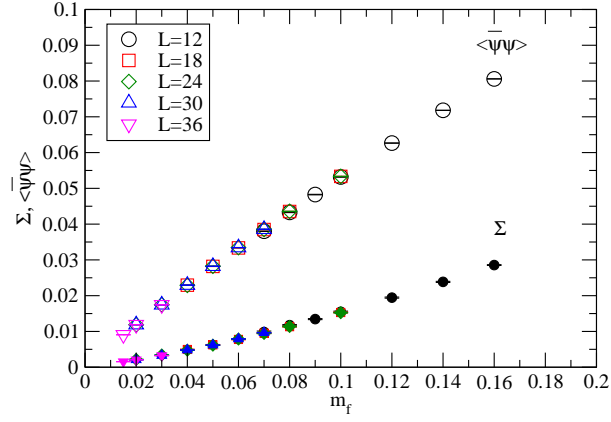}}
\caption{
$\langle {\bar \psi} \psi \rangle$ (Eq.~(\ref{eq:pbp})) and $\Sigma$
 (Eq.~(\ref{eq:Sigma})) as a function of $m_f$.
The open symbol represents $\langle {\bar \psi} \psi \rangle$
and the filled symbol is $\Sigma$.}
\label{fig:gmorpbp}
\end{figure}

\begin{figure}[!h]
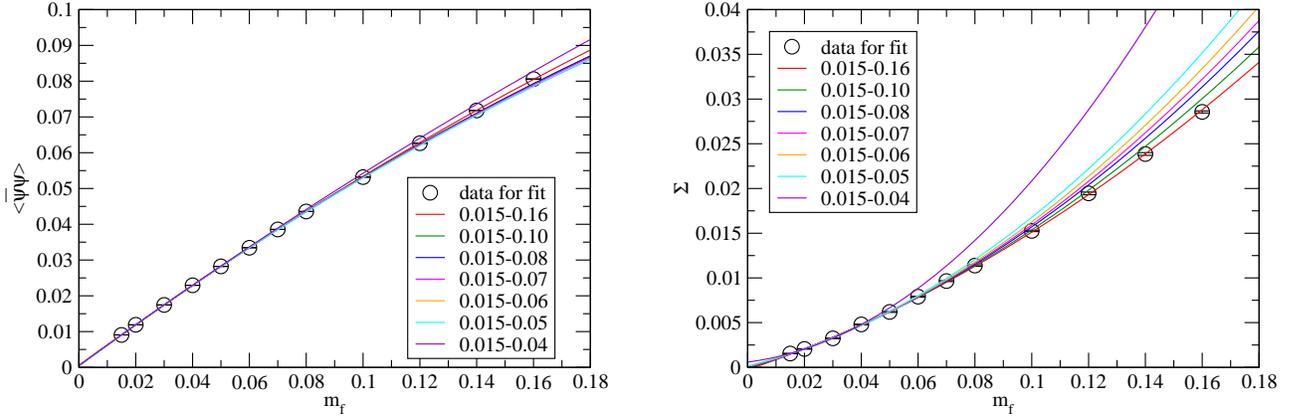
 
\makebox[.49\textwidth][r]{\includegraphics[scale=.45]{nf8_figures/fig-pbp-quad.eps}}
\makebox[.49\textwidth][r]{\includegraphics[scale=.45]{nf8_figures/fig-gmor-quad.eps}}
\caption{
Quadratic fits of $\langle {\bar \psi} \psi \rangle$ (left)
and $\Sigma$ (right).
}
\label{fig:pbpfit}
\end{figure}

\begin{figure}[!h]
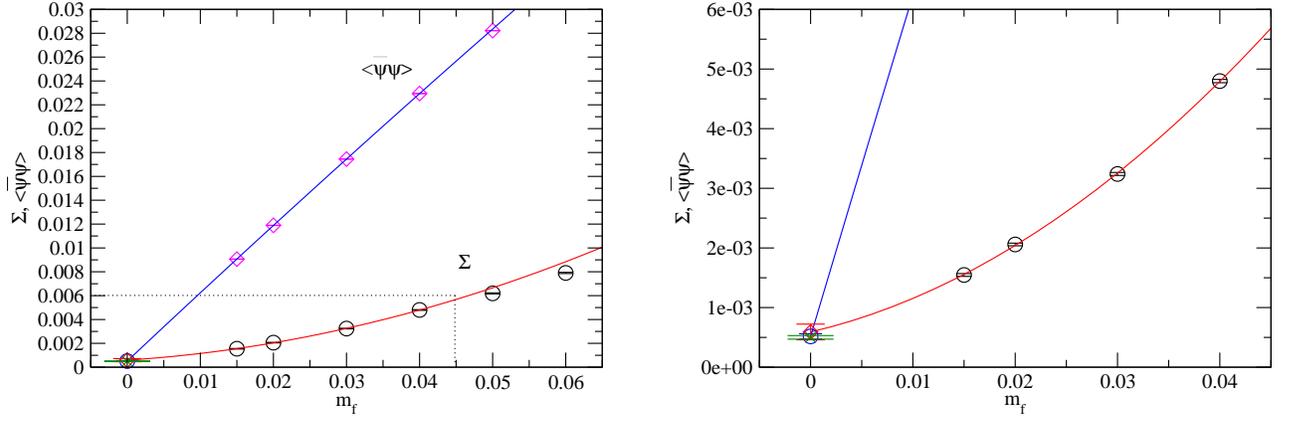
 
\makebox[.49\textwidth][r]{\includegraphics[scale=.45]{nf8_figures/fig-cond-climit-0.eps}}
\makebox[.49\textwidth][r]{\includegraphics[scale=.45]{nf8_figures/fig-cond-climit.eps}}
\caption{
$\Sigma$ and $\langle {\bar \psi} \psi \rangle$ (left panel) as a function of $m_f$.
The small region, $0 \leq m_f \leq 0.045$ and $0 \leq \Sigma, \langle {\bar \psi} \psi \rangle \leq 0.0006$,
 in the left panel is enlarged to the right panel.
The quadratic fit curves 
by using the data in $0.015 \leq m_f \leq 0.04$
are shown.
The green symbol is the value in Eq.~(\ref{eq:gmor2}).
}
\label{fig:cond}
\end{figure}

In this subsection,
we analyze the chiral condensate,
which is an order parameter of S$\chi$SB.
We perform a direct measurement of the chiral condensate
$\langle {\bar \psi} \psi \rangle = {\rm Tr} [D_{HISQ}^{-1}(x,x)]/4$
and compare  it with the quantity
\begin{equation}
 \Sigma \equiv \frac{F_\pi^2 M_\pi^2}{4 m_f},
  \label{eq:Sigma}
\end{equation}
which, in the chiral limit,  should coincide  with the chiral condensate
through the Gell-Mann-Oakes-Renner (GMOR) relation.
Fig.~\ref{fig:gmorpbp} shows the $\langle {\bar \psi} \psi \rangle$ and $\Sigma$
for each $m_f$.
%
We carry out the quadratic fits for each quantity,
whose results are summarized in Table~\ref{tab:condfit}
and shown in Fig.~\ref{fig:pbpfit}.
%
The chiral extrapolations for $\langle {\bar \psi} \psi \rangle$ and $\Sigma$
give good values of $\chi^2$/dof 
only in the small $m_f$ region $ 0.015 \leq m_f \leq 0.04$,
though the dof is too small.
Both the results in the chiral limit are non-zero,
and are consistent with each other, see Fig.~\ref{fig:cond}:
\begin{equation}
\left. \langle {\bar \psi} \psi \rangle \right|_{m_f \to 0} = 0.00052(5), \quad  \left. \Sigma \right|_{m_f \to 0}=0.00059(13) .
\label{eq:gmor1}
\end{equation}

We also estimate  the chiral condensate in the chiral limit 
by multiplying  $F$ in Eq.~\ref{eq:chiralF}
with the value of $M_\pi^2/m_f$ in the chiral limit obtained from linear fit
in Table~\ref{tab:condfit}:
\begin{equation}
F^2 \cdot \left. \left( \frac{M_\pi^2}{4 m_f} \right) \right|_{m_f \to 0} = 0.00050(3) ,
\label{eq:gmor2}
\end{equation}
which is consistent with those from the direct and indirect measurements.
%
\begin{table}
\caption{Chiral condensate in the chiral limit:
The quadratic fit result of $\langle {\bar \psi} \psi \rangle$  and $\Sigma$ in various fit ranges.
$\langle {\bar \psi} \psi \rangle=C_0^{\langle {\bar \psi} \psi \rangle}+C_1^{\langle {\bar \psi} \psi \rangle} m_f + C_2^{\langle {\bar \psi} \psi \rangle} m_f^2$.
$\Sigma=C_0^\Sigma+C_1^\Sigma m_f + C_2^\Sigma m_f^2$.
The linear fit of $M_\pi^2/m_f = C_0^{(M_\pi^2/m_f)} + C_1 m_f$
yields the combination $F^2 M_\pi^2/(4 m_f) \rightarrow  F^2 C_0^{(M_\pi^2/m_f)} / 4$ in the chiral limit.
}
\label{tab:condfit}
\begin{ruledtabular}
\begin{tabular}{cx{4,8}drx{4,8}drx{4,8}drx{4,8}}
\multicolumn{1}{c}{fit range} &
\multicolumn{3}{c}{$C_0^{\langle {\bar \psi} \psi \rangle}$} &
\multicolumn{3}{c}{$C_0^\Sigma$} &
\multicolumn{3}{c}{$C_0^{(M_\pi^2/m_f)}$} &
\multicolumn{1}{c}{$F^2 C_0^{(M_\pi^2/m_f)} / 4$} \\
\cline{2-4}\cline{5-7}\cline{8-10}
\multicolumn{1}{c}{($m_f$)}&
\multicolumn{1}{c}{value} &
\multicolumn{1}{c}{$\chi^2/\text{dof}$} &
\multicolumn{1}{c}{dof} &
\multicolumn{1}{c}{value} &
\multicolumn{1}{c}{$\chi^2/\text{dof}$} &
\multicolumn{1}{c}{dof} &
\multicolumn{1}{c}{value} &
\multicolumn{1}{c}{$\chi^2/\text{dof}$} &
\multicolumn{1}{c}{dof} &
\\
\hline
 0.015--0.04  &  0.00052(5)  &  2.65  &1&   0.00059(13) &  1.11  &1&  2.087(18)  &   1.34 &2&  0.00050(3)  \\
 0.015--0.05  &  0.00037(3)  &  11.5  &2&   0.00015(9)  & 11.5   & 2&  2.126(14)  &   5.22 &3&  0.00041(17) \\
 0.015--0.06  &  0.00037(2)  &  7.70  &3&   0.00007(7)  &  8.12  & 3&  2.151(11)  &   6.07 &4&  0.00043(13) \\
 0.015--0.07  &  0.00039(2)  &  6.51  &4&   0.00002(6)  &  6.56  & 4&  2.186(10)  &  11.4  &5&  0.00047(11) \\
 0.015--0.08  &  0.00041(2)  &  6.23  &5&  -0.00003(6)  &  6.22  & 5&  2.204(9)    &  14.2  &6&  0.00048(9)  \\
 0.015--0.10  &  0.00041(2)  &  5.19  &6&  -0.00013(4)  &  6.75  & 6&             &        &              \\
 0.015--0.16  &  0.00056(1)  &  18.9  &9&  -0.00026(3)  &  7.19  & 9&            &        &              \\
\end{tabular}
\end{ruledtabular}
\end{table}

From the analyses up to chiral log of all the observables,  
$F_\pi$, $M_\pi$, $M_\rho$ and the chiral condensate,
we find that chiral property of  $N_f=8$ QCD is consistent with that of S$\chi$SB.

\subsection{Chiral log corrections}
\label{subsec:log}

So far we have not included the logarithmic correction in the chiral fits.
Here we estimate such effects as systematic errors
on our previous results.

The logarithmic $m_f$ dependence is predicted by 
the next leading order (NLO) ChPT for 
both the $M_\pi^2/m_f$ and $F_\pi$~\cite{Gasser:1983yg}, 
whose formulae are given by
\begin{eqnarray}
\frac{M_\pi^2}{m_f} &=& 2B\left( 1 + \frac{x}{N_f}\log(x) + c_3 x\right)
\label{eq:chpt_mpi}
\\
F_\pi &=& F\left(1 - \frac{N_f\, x}{2}\log(x) + c_4 x\right),
\label{eq:chpt_fpi}
\end{eqnarray}
where the expansion parameter is denoted by $x = 4 B m_f / ( 4\pi F)^2$,
and $B, F, c_3$ and $c_4$ are the low energy constants.
Our data do not have such logarithmic dependence 
even in the lightest $m_f$ region
as shown in the previous subsections.
Actually, such a fit leads to a large $\chi^2$.
This is due to the fact that our $m_f$ is much heavier 
than  the region where the NLO ChPT is applicable.
The log correction of the $F_\pi$, 
however, is enhanced by the $N_f$ (Eq.~(\ref{eq:chpt_fpi})),
so that the $F$ might be largely affected 
by this correction, 
especially in this large $N_f$ theory.
Thus, we attempt to estimate the size of the correction 
by matching our polynomial fit results 
to the NLO ChPT at $m_f$ such that ${\cal X}=1$, with ${\cal X}$ defined
in Eq.~(\ref{eq:X}) 
where $F$ should read the re-estimated one in this analysis.
The details of the analysis are explained in Appendix~\ref{app:log}.
A reasonable value of the ${\cal X}\lesssim 1$ is realized
only in the region, $m_f \lesssim 0.002$,  much lighter 
than the $m_f$ used in our simulation.
From the analysis 
we find that the log correction reduces
the value of $F$ by about 30\% from the result with quadratic fit.

The other low energy constants including $B$ are obtained simultaneously.
The log correction of the chiral condensate is estimated 
from the GMOR relation,
$\left. \langle \overline{\psi} \psi \rangle\right|_{m_f\to 0} = BF^2/2$,
where the values of $F$ and $B$ estimated in this analysis are used.
We find that the chiral condensate is reduced 
by roughly half from the result with quadratic fit
by the log correction.

Apart from the log correction, we also estimate the systematic error
from other sources. While we adopted the quadratic chiral fits for $F$ and
$\left. \langle \overline{\psi} \psi \rangle\right|_{m_f\to 0}$,
linear fits work with reasonable $\chi^2$ with the same fitting range.
The differences are counted as systematic errors.
For the chiral condensate, the largest difference from the result
with the direct measurement to one of the indirect measurements is
counted as a systematic error. 

The results for the decay constant 
and the chiral condensate 
at the chiral limit in this work are
\begin{eqnarray}
F &=& 0.031(1)(^{+2}_{-10}),\\
\left. \langle \overline{\psi} \psi \rangle\right|_{m_f\to 0} &=&
0.00052(5)(^{+8}_{-29}),
\end{eqnarray}
where the first and second errors are 
statistical and systematic ones, respectively. 
The lower systematic errors are coming from the log corrections,
while the upper ones from the others.

It would be useful to estimate physical quantities 
in units of the $F$, 
because in the technicolor model
the $F$ is related to the weak scale,
\begin{equation}
\sqrt{N_d} F/\sqrt{2} = 246\ {\rm GeV},
\end{equation}
where $N_d$ is the number of the fermion weak doublets 
as $1 \le N_d \le N_f/2$.
From our result, 
the ratio $M_\rho/F$ in the chiral limit
is given as
\begin{equation}
\frac{M_\rho}{F/\sqrt{2}} = 7.7(1.5)(^{+3.8}_{-0.4}),
\label{eq:rho_f_ratio}
\end{equation}
where the $M_\rho$ in the chiral limit 
is the result of the quadratic fit in Eq.~({\ref{eq:M_rho}}).

In this analysis
we observe the large corrections 
of the chiral log term in ChPT.
In order to reduce the systematic error of the chiral extrapolation
and to obtain more accurate predictions in this theory,
we will need simulations at further small $m_f$ region on larger volumes.

\section{Study of remnants of conformality}
\label{sec:FSHS}

In the previous section 
we showed that the $N_f=8$ theory is in the  S$\chi$SB phase.
However, if this theory is near the conformal phase boundary,
it is expected that some remnants of the conformal symmetry appear
in physical quantities.

Here we start with an analysis of $F_\pi$ from a different point of view.
In the conformal phase the $F_\pi$ obeys 
the hyperscaling relation in the infinite volume, Eq.~(\ref{eq:hs}).
We perform the power fit
$F_\pi =  C_1 m_f^{1/(1+\gamma)}$ with various $m_f$ ranges, where
$C_1$ and $\gamma$ are free parameters.
The numerical results of the power fit are summarized in the
Table~\ref{tab:powerfit}. 
\begin{table}
\caption{Power fit results of $F_\pi$  for various fit ranges,
using $F_\pi= C_1 m_f^{1/(1+\gamma)}$. The left table shows the results for
the ranges with minimum mass set to the lightest, $m_f=0.015$, while
the right one does those with maximum mass being the heaviest $m_f=0.16$.
}
\label{tab:powerfit}
\begin{minipage}[t]{.47\textwidth}
\begin{ruledtabular}
\begin{tabular}{cx{3,6}x{3,6}d}
\multicolumn{1}{c}{fit range ($m_f$)} &
\multicolumn{1}{c}{$C_1$} &
\multicolumn{1}{c}{$\gamma$} &
\multicolumn{1}{c}{$\chi^2/{\rm dof}$} \\
\hline
0.015--0.04 & 0.415(7) & 0.988(19) & 14.8 \\
0.015--0.05 & 0.414(5) & 0.991(15) & 9.84 \\
0.015--0.06 & 0.418(4) & 0.979(12) & 7.88 \\
0.015--0.07 & 0.424(3) & 0.963(9) & 7.35 \\ 
0.015--0.08 & 0.425(3) & 0.961(8) & 6.15 \\ 
0.015--0.10 & 0.426(2) & 0.958(7) & 5.31 \\ 
0.015--0.16 & 0.428(1) & 0.952(4) & 3.98 \\ 
\end{tabular}
\end{ruledtabular}
\end{minipage}
\hfill
\begin{minipage}[t]{.47\textwidth}
\begin{ruledtabular}
\begin{tabular}{cx{3,6}x{3,6}d}
\multicolumn{1}{c}{fit range ($m_f$)} &
\multicolumn{1}{c}{$C_1$}  &
\multicolumn{1}{c}{$\gamma$} &
\multicolumn{1}{c}{$\chi^2/{\rm dof}$} \\
\hline 
0.02--0.16 & 0.429(1) & 0.947(4) & 2.22 \\ 
0.03--0.16 & 0.431(1) & 0.942(5) & 1.94 \\ 
0.04--0.16 & 0.429(2) & 0.950(10) & 1.23 \\
0.05--0.16 & 0.431(2) & 0.941(7) & 0.66 \\
0.06--0.16 & 0.429(2) & 0.948(9) & 0.44 \\
0.07--0.16 & 0.429(3) & 0.950(10) & 0.52 \\
0.08--0.16 & 0.431(3) & 0.939(14) & 0.20 \\
0.10--0.16 & 0.432(4) & 0.934(19) & 0.23 \\
\end{tabular}
\end{ruledtabular}
\end{minipage}
\end{table}

The power fit does not work in the lightest $m_f$ region,
$0.015 \le m_f \le 0.04$,
in which the $F_\pi$ is consistent with ChPT analysis 
and the $F$ is non-zero 
as presented in the previous section.
On the other hand, it is remarkable that 
the fit results in the mass range, $m_f \gtrsim  0.05$,
are consistent with the power behavior, 
the same way as the hyperscaling relation.
Furthermore the estimated $\gamma$ is stable in the larger mass region
(see the right table of Table \ref{tab:powerfit}), the property expected
from hyperscaling.
This suggests that, although $N_f=8$ QCD
is in the S$\chi$SB phase,
there exists a remnant of the conformality.
Therefore, in this section,
we will carry out further in depth analysis, which
employs the hyperscaling test on the finite volume for
$F_\pi$ as well as $M_\pi$ and $M_\rho$,
to investigate whether the remnant of the conformality 
really persists.

\subsection{Finite size hyperscaling test}
\label{subsec:fshs}

If the system is in the conformal window,
the data on a finite volume 
is in good agreement with the finite size hyperscaling (FSHS) 
having a universal value of $\gamma=\gamma_*$ 
at IRFP for observables as given in Eq.~(\ref{eq:FsHs}).
In general 
our data of $N_f=8$ cannot satisfy the FSHS 
with universal $\gamma$ in the whole range of $m_f$,
because we showed that the theory is in the S$\chi$SB phase
as analyzed in Sec.~\ref{sec:chpt}.
However, 
because of the power behavior in the middle range of the fermion mass
as mentioned in the above,
we carry out the FSHS test in our data
to find a remnant of the conformality.

\begin{figure}[!h]
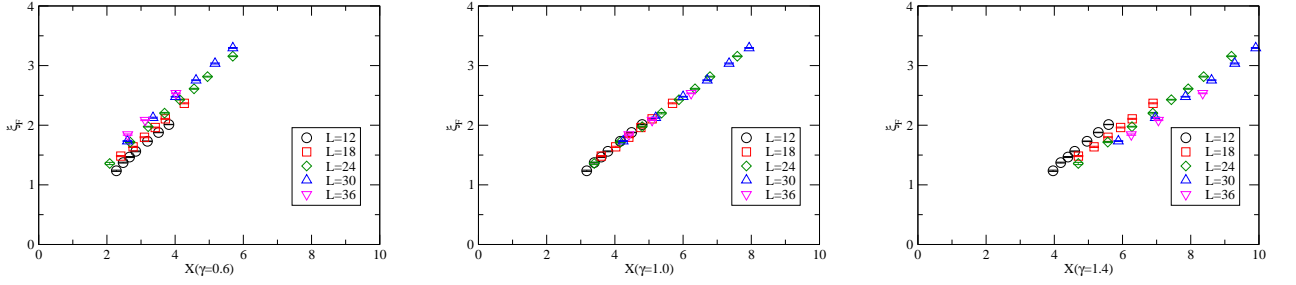
 
\makebox[.32\textwidth][r]{\includegraphics[scale=0.3]{nf8_figures/fig-hs-fpi-g0.6.eps}}
\makebox[.32\textwidth][r]{\includegraphics[scale=0.3]{nf8_figures/fig-hs-fpi-g1.0.eps}}
\makebox[.32\textwidth][r]{\includegraphics[scale=0.3]{nf8_figures/fig-hs-fpi-g1.4.eps}}
\caption{
$\xi_F$ plotted as functions of $X$ with $\gamma=0.6$ (left),
$1.0$ (center) and $1.4$ (right) for the FSHS test.
}
\label{fig:nf8fpihs}
\end{figure}

\begin{figure}[!h]
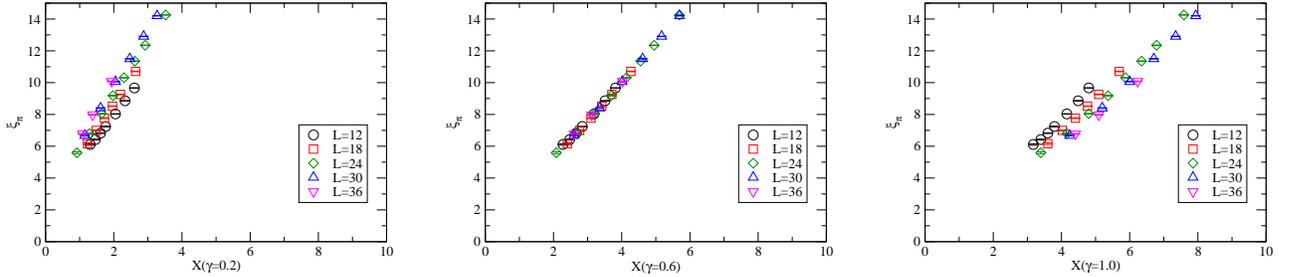
 
\makebox[.32\textwidth][r]{\includegraphics[scale=0.3]{nf8_figures/fig-hs-mpi-g0.2.eps}}
\makebox[.32\textwidth][r]{\includegraphics[scale=0.3]{nf8_figures/fig-hs-mpi-g0.6.eps}}
\makebox[.32\textwidth][r]{\includegraphics[scale=0.3]{nf8_figures/fig-hs-mpi-g1.0.eps}}
\caption{
$\xi_\pi$ plotted as functions of $X$ with $\gamma=0.2$ (left),
$0.6$ (center) and $1.0$ (right) for the FSHS test.
}
\label{fig:nf8mpihs}
\end{figure}

\begin{figure}[!h]
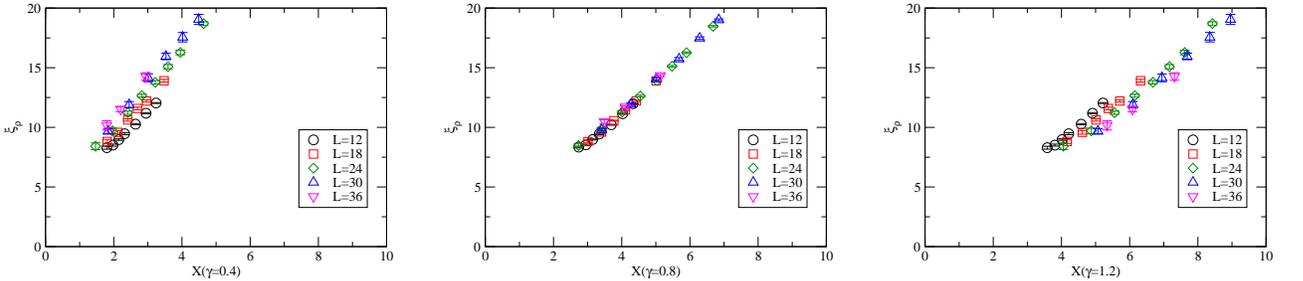
 
\makebox[.32\textwidth][r]{\includegraphics[scale=0.3]{nf8_figures/fig-hs-mpv-g0.4.eps}}
\makebox[.32\textwidth][r]{\includegraphics[scale=0.3]{nf8_figures/fig-hs-mpv-g0.8.eps}}
\makebox[.32\textwidth][r]{\includegraphics[scale=0.3]{nf8_figures/fig-hs-mpv-g1.2.eps}}
\caption{
$\xi_\rho$ plotted as functions of $X$ with $\gamma=0.4$ (left),
$0.8$ (center) and $1.2$ (right) for the FSHS test.
}
\label{fig:nf8mrhohs}
\end{figure}

For this test, 
we plot the observables, 
$\xi_F$ (Eq.~(\ref{eq:xiF})), $\xi_\pi$, and $\xi_\rho$ (Eq.~(\ref{eq:xip})), 
as functions of $X = L m_f^{1/(1+\gamma)}$
with changing the value of $\gamma$.
Figures~\ref{fig:nf8fpihs}, \ref{fig:nf8mpihs} and \ref{fig:nf8mrhohs}
are the results of the FSHS test of $\xi_F$, $\xi_\pi$ and $\xi_\rho$
for various $\gamma$'s:
The data are aligned (collapsing) 
at around $\gamma=1.0$, 0.6 and 0.8, respectively.
%
%
%
The optimal values of $\gamma$ for the  observables
are not universal in this estimate,
in contrast to $N_f=12$, where the alignments was
observed with almost universal $\gamma$ \cite{Aoki:2012eq}.
It is also noted the existence of alignment for each observable
is in contrast to $N_f=4$, where no alignment is observed
(see Appendix~\ref{app:nf4}).

Since Figs.~\ref{fig:nf8fpihs}, \ref{fig:nf8mpihs} and \ref{fig:nf8mrhohs},
show the good linear behavior,
we carry out a linear fit as the leading approximation of FSHS,
\begin{equation}
\xi_H = C_0^H + C_1^H X,
\label{eq:FSHF_linear}
\end{equation}
for each observable.
This formula becomes the hyperscaling, Eq.~(\ref{eq:hs})
in the infinite volume limit.
In Sec.~\ref{sec:chpt} we saw the ChPT fit worked well 
for the smallest mass region $m_f \le 0.04$ for all the observables.
On  the other hand, we already showed the power-like behavior of
the $F_\pi$ for larger masses $m_f\ge 0.05$.
Thus, we restrict ourselves
to fit the data in $m_f \ge 0.05$ in this analysis.
To have good linearity we restrict 
the data in the larger $\xi_\pi$ region, $\xi_\pi \ge 8$.

\begin{figure}[!h]
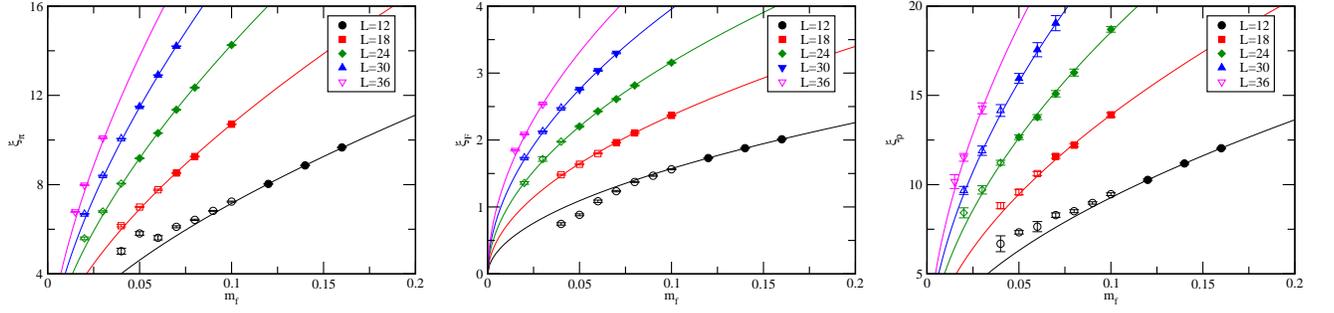
 
\makebox[.32\textwidth][r]{\includegraphics[scale=0.3]{nf8_figures/fshsfit-mpi.eps}}
\makebox[.32\textwidth][r]{\includegraphics[scale=0.3]{nf8_figures/fshsfit-fpi.eps}}
\makebox[.32\textwidth][r]{\includegraphics[scale=0.3]{nf8_figures/fshsfit-mpv.eps}}
\caption{
Linear fits for the FSHS of $M_\pi$ (left), $F_\pi$ (center), and
$M_\rho$ (right).
The filled symbols are included in the fit, 
but the open symbols are omitted.
The fitted region is $m_f \ge 0.05$ and $\xi_\pi \geq 8$.
}
\label{fig:nf8fpihsfit}
\end{figure}

Panels in Fig.~\ref{fig:nf8fpihsfit} 
are the fit result of FSHS 
for $\xi_\pi$, $\xi_F$ and $\xi_\rho$
from the left to the right.
The fitting result is given in Table~\ref{tab:fshsfit},
which is consistent with the $P(\gamma)$ analysis 
that does not assume the functional form of fitting
given in the Appendix~\ref{app:pgamma}.
%
\begin{table}
\caption{The $\gamma$ fitted by the linear ansatz.
The fitted region is $m_f \ge 0.05$ and $\xi_\pi \geq 8$.}
\label{tab:fshsfit}
\begin{minipage}[t]{.7\textwidth}
\begin{ruledtabular}
\begin{tabular}{cdddd}
&
\multicolumn{1}{c}{$\gamma$} &
\multicolumn{1}{c}{$C_0^H$} &
\multicolumn{1}{c}{$C_1^H$} &
\multicolumn{1}{c}{$\chi^2/\text{dof}$} \\
\hline
$\xi_\pi$  & 0.5668(26) &   0.049(22) & 2.57766(99) & 2.52 \\
$\xi_F$    & 0.9279(79) &  -0.17(10)  & 0.4372(38)  & 0.73 \\
$\xi_\rho$ & 0.798(20)  &   0.04(19)  & 2.779(69)   & 0.66 \\
\end{tabular}
\end{ruledtabular}
\end{minipage}
\end{table}
In Fig.~\ref{fig:nf8fpihsfit} data included in the fit are shown as
filled symbols, while those with open symbols are not included.
The figures show the linear fit works well for $\xi_\rho$ 
($\chi^2/{\rm dof} =0.66$) and $\xi_F$ ($\chi^2/{\rm dof}=0.73$) ,
while the fit for $\xi_\pi$ has $\chi^2/{\rm dof}=2.52$.

The larger $\chi^2/$dof of the $\xi_\pi$ fit 
might be caused by corrections 
which are not explained by the simple fit 
form in Eq.(\ref{eq:FSHF_linear}).
To check the existence of the correction, 
we fit the data of $\xi_\pi$ only on two volumes, 
and slide the range of the volumes 
to investigate the  fit range dependence of the $\gamma$.
The resulting $\chi^2/$dof,
tabulated in Table~\ref{tab:fshsfit_vol},
is better than the above fit.
The results seem to have a tendency to decease the $\gamma$ 
as the volume decreased.
The maximum and minimum results deviate from each other 
by more than two standard deviations,
and they also differ from the result using the four volumes
tabulated in Table~\ref{tab:fshsfit}.
This tendency would suggest 
that there are corrections
to the leading behavior of the FSHS in Eq.(\ref{eq:FSHF_linear}),
but it is not clear
that this tendency comes from only a finite volume effect, 
because the range of the $m_f$ 
is also changed as the volume.
On the other hand, 
the results for the $\xi_F$ and $\xi_\rho$ in Table~\ref{tab:fshsfit_vol}
are consistent with each result tabulated in Table~\ref{tab:fshsfit},
so that we do not expect 
that there are significant corrections
in these data.

We will discuss 
the types of
the corrections of Eq.(\ref{eq:FSHF_linear})
in the next subsection.

\begin{table}
\caption{The $\gamma$ fitted
by the linear ansatz using the data on two volumes.
}
\label{tab:fshsfit_vol}
\begin{ruledtabular}
\begin{tabular}{cx{2,4}dx{2,4}dx{2,4}d}
&
\multicolumn{2}{c}{$\xi_\pi$} &
\multicolumn{2}{c}{$\xi_F$} &
\multicolumn{2}{c}{$\xi_\rho$} \\
\cline{2-3}\cline{4-5}\cline{6-7}
\multicolumn{1}{c}{$L$} &
\multicolumn{1}{c}{$\gamma$} &
\multicolumn{1}{c}{$\chi^2/\text{dof}$} &
\multicolumn{1}{c}{$\gamma$} &
\multicolumn{1}{c}{$\chi^2/\text{dof}$} &
\multicolumn{1}{c}{$\gamma$} &
\multicolumn{1}{c}{$\chi^2/\text{dof}$} \\
\hline
(30,24) & 0.5864(61) & 1.09 & 0.948(18) & 1.19 & 0.84(11)  & 0.62 \\
(24,18) & 0.5720(88) & 0.66 & 0.934(23) & 0.88 & 0.765(58) & 1.25 \\
(18,12) & 0.5509(54) & 0.49 & 0.924(12) & 0.15 & 0.809(28) & 1.00 \\
\end{tabular}
\end{ruledtabular}
\end{table}

\subsection{FSHS fits with the correction term}
\label{subsec:gfit}

Since $N_f=8$ theory is in S$\chi$SB phase,
FSHS cannot become accurate 
by approaching to the chiral limit, 
which is in contrast to the $N_f=12$
where FSHS does \cite{Aoki:2012eq}. 
Therefore FSHS is only expected for larger mass region, 
where mass corrections may not be negligible.
In fact in the last subsection 
the decreasing tendency of the $\gamma(M_\pi)$
depending on the fit range is seen, 
which might suggest that there are corrections in the
simple FSHS form in Eq.(\ref{eq:FSHF_linear}),
To include mass corrections
we assume the same fitting forms
as in the $N_f=12$ case~\cite{Aoki:2012eq} as,
\begin{equation}
\xi_H = C^H_0 + C^H_1 X + C^H_2 Lm_f^\alpha.
\label{eq:FSHS_correction}
\end{equation}
Since it is hard to determine the exponent $\alpha$ of the correction term
when the fit is performed for each observable individually,
we fix it in our analysis.
Among various choices of the $\alpha$, 
we take two values: $\alpha=1$ and $2$.
The first choice $\alpha=1$ is regarded as an $m_f$ correction
in the heavy region,  
and the second one $\alpha=2$ may be identified 
as a $\mathcal{O}(a^2)$ discretization effect.

Using the fit assumptions 
we fit each observable
 with the same data region 
 as in the last subsection, 
 $m_f \ge 0.05$ and $\xi_\pi \ge 8$.
The results are tabulated in Table~\ref{tab:FSHS_correction}.
The fit results with both $\alpha=1$ and 2 of the $\xi_\pi$ show 
the correction term actually takes effect ($C_2^\pi\ne 0$),
with reasonable $\chi^2/{\rm dof}$.
Due to the large correction, 
the $\gamma$ of the $\xi_\pi$ is largely changed
from the one without the correction term in Table~\ref{tab:fshsfit},
especially in the $\alpha = 1$ case, 
and the value becomes closer
to the ones from the other observables.
On the other hand, 
for the $\xi_F$ and $\xi_\rho$ fits, 
it is found that 
the correction is negligible,
and the resulting $\gamma$'s are consistent with the ones 
without the correction, 
presented in Table~\ref{tab:fshsfit},
as expected in the analyses in the last subsection.
While in the $\alpha = 1$ case, 
we obtain reasonable consistency of 
the $\gamma$ from the three observables 
within less than two standard deviations,
we cannot exclude the $\alpha = 2$ fit.
Thus, the above analyses would suggest $\gamma = 0.62$--0.97
depending on the observables and also the form of the correction term.

\begin{figure}[!b]
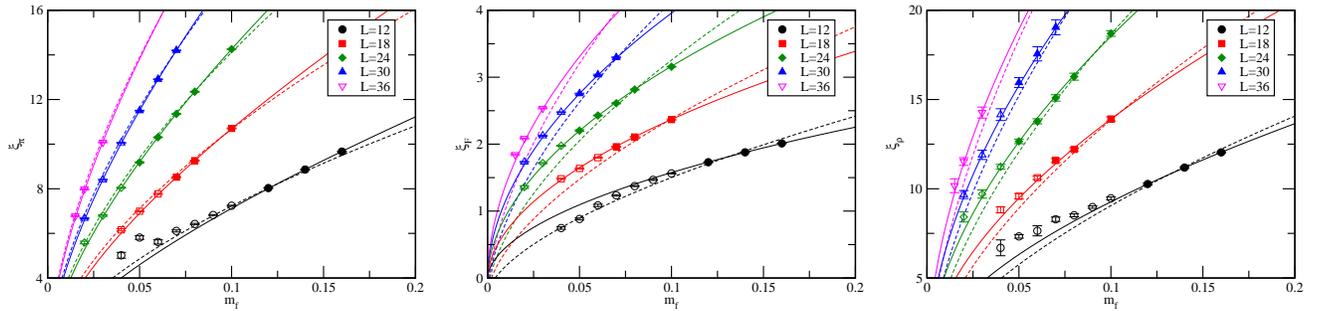
 
\makebox[.32\textwidth][r]{\includegraphics[scale=0.30]{nf8_figures/mpiL.eps}}
\makebox[.32\textwidth][r]{\includegraphics[scale=0.30]{nf8_figures/fpiL.eps}}
\makebox[.32\textwidth][r]{\includegraphics[scale=0.30]{nf8_figures/rhoL.eps}}
\caption{
Simultaneous FSHS fit in $\xi_\pi$(left), $\xi_F$(center) and $\xi_\rho$(right)
with $\alpha = 1$.
The filled symbols are included in the fit,
but the open symbols are omitted.
The fitted region is $m_f \ge 0.05$ and $\xi_\pi \geq 8$.
The solid curve is the fit result.
For a comparison, the simultaneous fit result without correction terms
is also plotted by the dashed curve, whose $\chi^2/dof$ = 83.
}
\label{fig:nf8-gFSHS}
\end{figure}

Since we observed that the values of $\gamma$ with
Eq.~(\ref{eq:FSHS_correction}) for all the observables 
become closer to each other than those
without the correction terms,
it might be possible 
to obtain a common value of the $\gamma$ 
from all the observables 
using the fit including the correction.
Thus, we perform simultaneous fit 
using all the observables 
$M_\pi$, $F_\pi$, and $M_\rho$ with a common $\gamma$. 
For simplicity, 
we assume the absence of the statistical correlations
between each data of $M_\pi$, $F_\pi$, and $M_\rho$.
In the fit we do not fix the value of the $\alpha$, 
and treat it as a free parameter.
It is expected 
that the corrections are small 
in the $\xi_F$ and $\xi_\rho$,
so that we first carry out a fit 
omitting the correction term in the $\xi_F$.
The result is summarized in Table~\ref{tab:FSHS_correction_global_fpi}.
This fit works well, 
and gives a reasonable value of the $\chi^2/$dof.
The resulting $\alpha$ is close to unity.
Similar value of $\alpha$ is also obtained from a fit 
without  the correction term in the $\xi_\rho$
as shown in Table~\ref{tab:FSHS_correction_global_rho}.
This means 
that the exponent of the correction term 
is close to unity in our data, while $\gamma$'s from the two fits
are different each other.
The difference is regarded as the ambiguity in this estimate.
It is also possible to carry out a simultaneous fit 
without the corrections 
in both the $\xi_F$ and $\xi_\rho$, 
and fits with the correction terms for all the observables
using the fixed $\alpha = 1$ and $\alpha=(3-2\gamma)/(1+\gamma)$,
because our data prefer $\alpha \sim 1$ in the above fits.
Note that the last one is 
inspired by the analytic expression of 
the solution of the Schwinger-Dyson equation~\cite{Aoki:2012ve}.
Fig.~\ref{fig:nf8-gFSHS}
shows the fit result with $\alpha = 1$ 
as a typical result of the simultaneous fit.
These results are shown in Table~\ref{tab:FSHS_correction_global_fpi_rho}
and their $\gamma$'s agree within the above ambiguity.
Under the assumption 
that all the observables give a universal $\gamma$,
we estimate $\gamma = 0.78$--0.93.

It is noted that a simultaneous fit including the lighter mass with 
$m_f\ge 0.015$ in $\xi_\pi \ge 6.8$ fails with a large $\chi^2$/dof$ =
3.5$ even if the mass correction is included.
This is because the chiral property is dictated by S$\chi$SB and should
not be consistent with universal hyperscaling near the chiral limit.

To summarize,
using the fits with the correction term, 
we estimated the value of $\gamma$ 
from the three observables, 
and obtain $\gamma = 0.62$--0.97
which depends on the observables 
and the correction term in the fit form.
Furthermore we carry out simultaneous FSHS fits 
with the correction term,
since a universal $\gamma$ would be expected 
if the theory is very close to the conformal phase boundary 
even in the S$\chi$SB phase.
The resulting $\gamma$ in the simultaneous fits reads 0.78--0.93.
These estimated $\gamma$'s would be identified as 
the  mass anomalous dimension 
in the walking regime.
\begin{table}
\caption{FSHS fit with a correction term.
The fit function: $\xi = C_0^H + C_1^H X + C_2^H Lm_f^\alpha$.
The fitted region is $m_f \ge 0.05$ and $\xi_\pi \geq 8$.}
\label{tab:FSHS_correction}
\begin{minipage}[t]{.7\textwidth}
\begin{ruledtabular}
\begin{tabular}{cx{2,4}x{2,4}x{2,4}x{2,4}x{2,4}}
$\alpha=1$ &
\multicolumn{1}{c}{$\gamma$} &
\multicolumn{1}{c}{$C_0^H$} &
\multicolumn{1}{c}{$C_1^H$} &
\multicolumn{1}{c}{$C_2^H$} &
\multicolumn{1}{c}{$\chi^2/\text{dof}$} \\
\hline 
$\xi_\pi$  
& 0.791(57) & -0.004(25)  & 1.74(14)  &  1.12(20)    & 0.66 \\
$\xi_F$ 
& 0.965(91) & -0.016(11)  & 0.419(44) &  0.026(65)   & 0.74 \\ 
$\xi_\rho$ 
& 0.80(25)  &  0.003(190) & 2.78(99)  & -0.01(1.30)  & 0.73 \\
\end{tabular}
\end{ruledtabular}
\end{minipage}
\begin{minipage}[t]{.7\textwidth}
\begin{ruledtabular}
\begin{tabular}{cx{2,4}x{2,4}x{2,4}x{2,4}x{2,4}}
$\alpha=2$ &
\multicolumn{1}{c}{$\gamma$} &
\multicolumn{1}{c}{$C_0^H$} &
\multicolumn{1}{c}{$C_1^H$} &
\multicolumn{1}{c}{$C_2^H$} &
\multicolumn{1}{c}{$\chi^2/\text{dof}$} \\
\hline 
$\xi_\pi$  
& 0.620(12) & -0.001(25)  & 2.421(35) &  0.98(21)  & 0.74 \\
$\xi_F$ 
& 0.941(30) & -0.016(10)  & 0.432(11) &  0.030(72) & 0.74 \\ 
$\xi_\rho$ 
& 0.792(83) &  0.001(190) & 2.80(23)  & -0.1(1.2)  & 0.73 \\
\end{tabular}
\end{ruledtabular}
\end{minipage}
\end{table}

\begin{table}
\caption{Simultaneous FSHS fit with a correction term,
$\xi = C_0^H + C_1^H X + C_2^H Lm_f^\alpha$,
where $\alpha$ is free parameter, but $C_2^F = 0$.
The fitted region is $m_f \ge 0.05$ and $\xi_\pi \geq 8$.
Degrees of freedom equals to 32.}
\label{tab:FSHS_correction_global_fpi}
\begin{minipage}[t]{.6\textwidth}
\begin{ruledtabular}
\begin{tabular}{x{2,4}x{2,4}x{2,4}}
\multicolumn{1}{c}{$\gamma$} &
\multicolumn{1}{c}{$\alpha$} &
\multicolumn{1}{c}{$\chi^2/\text{dof}$} \\
\hline
0.9292(82) & 0.879(53) & 0.67 \\
\end{tabular}
\end{ruledtabular}
\end{minipage}
\begin{minipage}[t]{.6\textwidth}
\begin{ruledtabular}
\begin{tabular}{cddd}
&
\multicolumn{1}{c}{$C_0^H$} &
\multicolumn{1}{c}{$C_1^H$} &
\multicolumn{1}{c}{$C_2^H$} \\
\hline 
$\xi_\pi$  
&  -0.004(25) & 1.290(98)  & 1.538(39) \\
$\xi_F$ 
&  -0.015(10) & 0.4366(38) & \multicolumn{1}{c}{---} \\ 
$\xi_\rho$ 
&   0.01(19)  & 2.280(64)  & 0.61(10) \\
\end{tabular}
\end{ruledtabular}
\end{minipage}
\end{table}

\begin{table}
\caption{Simultaneous FSHS fit with a correction term,
$\xi = C_0^H + C_1^H X + C_2^H Lm_f^\alpha$,
where $\alpha$ is free parameter, but $C_2^\rho = 0$.
The fitted region is $m_f \ge 0.05$ and $\xi_\pi \geq 8$.
Degrees of freedom equals to 32.}
\label{tab:FSHS_correction_global_rho}
\begin{minipage}[t]{.6\textwidth}
\begin{ruledtabular}
\begin{tabular}{x{2,4}x{2,4}x{2,4}}
\multicolumn{1}{c}{$\gamma$} &
\multicolumn{1}{c}{$\alpha$} &
\multicolumn{1}{c}{$\chi^2/\text{dof}$} \\
\hline
0.807(18) & 0.949(74) & 0.77 \\
\end{tabular}
\end{ruledtabular}
\end{minipage}
\begin{minipage}[t]{.6\textwidth}
\begin{ruledtabular}
\begin{tabular}{cddd}
&
\multicolumn{1}{c}{$C_0^H$} &
\multicolumn{1}{c}{$C_1^H$} &
\multicolumn{1}{c}{$C_2^H$} \\
\hline 
$\xi_\pi$  
&  -0.001(25) & 1.65(11)  & 1.174(66) \\
$\xi_F$ 
&  -0.011(10) & 0.516(15) & -0.107(19) \\ 
$\xi_\rho$ 
&   0.04(18)  & 2.754(63) & \multicolumn{1}{c}{---} \\
\end{tabular}
\end{ruledtabular}
\end{minipage}
\end{table}

\begin{table}
\caption{Simultaneous FSHS fit with a correction term,
$\xi = C_0^H + C_1^H X + C_2^H Lm_f^\alpha$ using
several choices of $\alpha$.
The fitted region is $m_f \ge 0.05$ and $\xi_\pi \geq 8$.}
\label{tab:FSHS_correction_global_fpi_rho}
\begin{minipage}[t]{.7\textwidth}
\begin{ruledtabular}
\begin{tabular}{cx{1,4}x{1,4}x{1,4}}
$\alpha=0.889(55)$ &
\multicolumn{1}{c}{$C_0^H$} &
\multicolumn{1}{c}{$C_1^H$} &
\multicolumn{1}{c}{$C_2^H$} \\
\hline 
$\xi_\pi$  & -0.005(25)  & 1.338(96)  & 1.494(37) \\
$\xi_F$    & -0.0275(98) & 0.4435(36) & \multicolumn{1}{c}{---} \\ 
$\xi_\rho$ &  0.53(16)   & 2.476(39)  & \multicolumn{1}{c}{---} \\
\hline
\multicolumn{4}{c}{
$\gamma=0.9130(76)$, $\chi^2/\text{dof}=1.73$, $\text{dof}=33$}
\end{tabular}
\end{ruledtabular}
\end{minipage}
\begin{minipage}[t]{.7\textwidth}
\begin{ruledtabular}
\begin{tabular}{cx{1,4}x{1,4}x{1,4}}
$\alpha=1$ fixed &
\multicolumn{1}{c}{$C_0^H$} &
\multicolumn{1}{c}{$C_1^H$} &
\multicolumn{1}{c}{$C_2^H$} \\
\hline 
$\xi_\pi$  & -0.014(24) & 1.61(10)  & 1.31(15)   \\
$\xi_F$    & -0.012(10) & 0.484(30) & -0.068(44) \\ 
$\xi_\rho$ &  0.01(19)  & 2.60(17)  & 0.25(24)   \\
\hline
\multicolumn{4}{c}{
$\gamma=0.874(25)$, $\chi^2/\text{dof}=0.75$, $\text{dof}=32$}
\end{tabular}
\end{ruledtabular}
\end{minipage}
\begin{minipage}[t]{.7\textwidth}
\begin{ruledtabular}
\begin{tabular}{cx{1,4}x{1,4}x{1,4}}
$\alpha=\frac{3-2\gamma}{1+\gamma}$ fixed &
\multicolumn{1}{c}{$C_0^H$} &
\multicolumn{1}{c}{$C_1^H$} &
\multicolumn{1}{c}{$C_2^H$} \\
\hline 
$\xi_\pi$  &  0.020(24) & 1.52(39)  &  1.17(35)  \\ 
$\xi_F$    & -0.011(10) & 0.572(34) & -0.158(52) \\ 
$\xi_\rho$ &  0.03(19)  & 2.91(30)  & -0.15(36)  \\
\hline
\multicolumn{4}{c}{
$\gamma=0.775(56)$, $\chi^2/\text{dof}=0.93$, $\text{dof}=32$}
\end{tabular}
\end{ruledtabular}
\end{minipage}
\end{table}
%

%
%

\section{Summary and Discussion}
\label{sec:summary}
In search for a candidate for the Walking Technicolor, 
we have investigated meson spectrum of $N_f=8$ QCD 
by the lattice simulations  based on the HISQ action 
for $\beta=6/g^2 =3.8$,
and for the fermion bare mass range $m_f=0.  015 - 0.16$ 
depending on the volume size $L^3 \times T$ 
with $(L,T)=(12,16),(18,24),(24,32),(30,40),(36,48)$.  
 
We found 
that the data of $F_\pi$, $M_\pi$ are consistent with the S$\chi$SB
well described by the ChPT, 
suggesting that $F\equiv F_\pi(m_f \rightarrow 0) \ne 0$,  and $M_\pi(m_f\rightarrow 0) =0$,  and $M_\rho$ 
is also non-vanishing $M_\rho(m_f\rightarrow 0)\ne 0$
in the chiral limit extrapolation. 
the $\rho$ mass in units of the $\pi$ decay constant was determined 
and shown in Eq.~(\ref{eq:rho_f_ratio}).
We further found that the chiral condensate $\langle \bar \psi \psi\rangle$ 
also has a non-zero value in the chiral limit, which nicely coincides
with those from the GMOR relation in that limit.
In these analysis we used $0.015\le m_f\le 0.04$.

The salient feature of our collaboration is 
that we have been investigating $N_f= 4, 8, 12, 16$ 
on the setting of HISQ action with the same systematics
in order to study the $N_f$-dependence of the physics systematically
\cite{Aoki:2012kr,Aoki:2012eq,Aoki:2012ep,Aoki:2013dz}. 
Thus our analyses for $N_f=8$ are made in comparison 
with those for $N_f=4$ and $12$ of our group.
The qualitative features of $N_f=8$ near the chiral limit 
were found to be similar to those of 
the $N_f=4$ case: The $N_f=4$ data indicated robust signals of S$\chi$SB phase.
The result was contrasted with that of the $N_f=12$ 
where in previous study 
the ChPT analysis was not self-consistent, 
while the finite-size hyperscaling (FSHS) relation held consistently 
with the conformal window.

We then checked whether this S$\chi$SB phase 
is close to the conformal window, 
having some remnant of the infrared conformality.
Remarkably enough, 
in contrast to the data  near the chiral limit ($m_f\le 0.04$)
indicating  the S$\chi$SB,  
those for  the relatively large fermion bare mass $m_f\ge 0.05$ 
away from the chiral limit 
actually exhibited a FSHS
with the scaling exponent 
$\gamma(M_\pi) \simeq 0.57$ ($\chi^2/{\rm dof}=2.5$), 
$\gamma(F_\pi) \simeq 0.93$ ($\chi^2/{\rm dof}=0.7$),
$\gamma(M_\rho) \simeq 0.80$ ($\chi^2/{\rm dof}=0.7$).
The value of $\gamma$ is non-universal depending on the observable,
with $\chi^2/{\rm dof}$ for $M_\pi$ being large compared to the others.
The existence of FSHS is in contrast to our $N_f=4$ data 
which we showed no trace of the hyperscaling even for large $m_f$ region
and hence no sign of the conformality as in the ordinary QCD.
This implies that there exists a remnant of the infrared conformality
where the S$\chi$SB effects are negligible (for 
a schematic view, see Case 2 in Fig.~\ref{fig:walking}). 

These hyperscaling relations were obtained 
for relatively large $m_f$, $m_f\ge 0.05$,
in contrast to the FSHS in conformal window, where
FSHS becomes arbitrarily accurate in the chiral limit.
Therefore, there could exist large mass corrections
 on the hyperscaling relation. 
If we include possible mass corrections on FSHS for each observable,
we obtained $\gamma$, $0.62 \lesssim \gamma \lesssim 0.97$.
Here the $\chi^2$/dof for $M_\pi$ was improved to the level of the others.

We  then performed a simultaneous fit over all the observables based on
certain model fitting functions including the mass corrections
to see if the universality of $\gamma$ can be improved by the corrections.
To our surprise 
we found that 
certain model fitting functions in fact yield a universal value 
of $\gamma$,  $0.78 \lesssim \gamma \lesssim 0.93$, 
with the variations depending on the model fitting function.
Since this result is obtained at a single value of  $\beta$,
it is important to see this feature holds in the continuum limit,
which will be studied by carrying out simulations with multiple values of $\beta$.
Possibility of such a large scaling exponent $\gamma$ was discussed
using Dirac eigen modes in Ref.~\cite{Cheng:2013eu}.

The anomalous dimension discussed in the walking technicolor 
is of course the value in the chiral limit. 
Lesson from the SD equation analysis in the conformal window 
tells us  \cite{Aoki:2012ve} that the value of $\gamma$ obtained by
the hyperscaling relation without corrections is an ``effective'' one
which is distorted by the mass corrections.
On the other hand,
the value determined by the fit explicitly incorporating the
mass corrections  
just corresponds to the $\gamma_*$ ($\gamma_m$ at the infrared fixed
point) in the chiral limit and  
hence is of direct relevance to the walking technicolor. 
Although the $\gamma_m$ so determined is the value at infrared scale, 
it coincides with
the one discussed in the walking technicolor 
evaluated at the ultraviolet scale,
as far as  the infrared conformality for the wide scale hierarchy exists.

Finally, 
we should comment on the possible light  flavor-singlet scalar meson 
in $N_f=8$ QCD. 
The walking technicolor predicts~\cite{Yamawaki:1985zg,Bando:1986bg} 
a  light composite Higgs-like scalar boson, 
the techni-dilaton,
as a pseudo NG boson of the approximate scale invariance 
inherent to the walking dynamics. Actually, 
it was shown~\cite{Matsuzaki:2012mk,Matsuzaki:2012xx}  
that the techni-dilaton is consistent with all the current data of the 125 GeV boson 
discovered at LHC.
Then, if the $N_f=8$ QCD behaves as a walking theory 
with approximate scale invariance, 
it would be expected that a light flavor-singlet scalar composite does exist. 
These studies are currently
under way and details will be reported elsewhere. 
Since the quantum number of such an object 
is the same as that of the scalar glueballs 
which may also be light, 
the lattice analyses near infrared conformality should be done 
with great care about the possible mixing with each other.
As such we made preliminary studies of both flavor-singlet scalar and
scalar glueballs for $N_f=12$.
We found a hint of a flavor-singlet scalar bound state lighter than $\pi$ for $N_f=12$~\cite{Aoki:2013pca}.
 
Summarizing all our analyses we may infer that a typical technicolor (``one-family model'')  
as modeled by the $N_f=8$ QCD can be a walking technicolor theory 
having an approximate scale invariance with large anomalous dimension
$\gamma_m \sim 1$.

\acknowledgments
We would like to thank Anna Hasenfratz, Yoichi Iwasaki and Julius Kuti for fruitful discussions,
and acknowledge Katsuya Hasebe for encouragements.
We also thank Enrico Rinaldi for helpful discussions.
Numerical simulation has been carried out on 
the supercomputer system $\varphi$ at KMI, Nagoya University
and on the computer facilities of the Research Institute for Information
Technology at Kyushu University.
This work is supported
by the JSPS Grant-in-Aid for Scientific Research (S) No.22224003, 
 (C) No.23540300 (K.Y.) and (C) No.21540289 (Y.A.),
and also by Grants-in-Aid of the Japanese Ministry for Scientific Research 
on Innovative Areas No. 23105708 (T.Y.).

\appendix

\section{Data tables in $N_f=8$ case}
\label{app:data}

The $N_f=8$ simulations are done at $\beta=3.8$ with the fixed aspect
ratio $L/T=3/4$: $(L, T)=(12, 16), (18, 24), (30, 40)$ and $(36, 48)$
and with various fermion masses $m_f$.
Resultant values of $F_\pi$, $M_\pi$, $M_{SC}$, $M_{\rho(PV)}$,
$M_{\rho(VT)}$ and $\langle\bar{\psi} \psi\rangle$ for each parameter,
together with the number of trajectories $N_{\rm{trj}}$ 
used for the measurements after thermalzation
are summarized in Tables~\ref{tab:1}, ~\ref{tab:2},  ~\ref{tab:3},
~\ref{tab:4} and ~\ref{tab:5}. 
\begin{table}
\caption{Results of the spectra on $V=12^3 \times 16$,
with $t_{\min}=10$.}
\label{tab:1}
\squeezetable
\begin{ruledtabular}
\begin{tabular}{lrllllll}
$m_f$ & $N_{\rm{trj}}$ & $F_\pi$ & $M_\pi$ & $M_{SC}$ & $M_{\rho(PV)}$ & $M_{\rho(VT)}$ & $\langle {\bar \psi} \psi \rangle$ \\ 
\hline
0.04 & 1224 & 0.0622(15) & 0.4181(110)& 0.4397(113) & 0.5574(370)   & 0.5389(360) & 0.02167(4) \\
0.05 & 1284 & 0.0735(12) & 0.4844(65)  & 0.4987(72)    & 0.6108(81)     & 0.6157(71)   & 0.02704(5) \\
0.06 & 1224 & 0.0904(15) & 0.4681(79)  & 0.4664(120)  & 0.6372(237)   & 0.6343(218)& 0.03269(6) \\
0.07 & 1264 & 0.1030(9)   & 0.5091(38)  & 0.5168(64)     & 0.6910(102)   & 0.6882(87)  & 0.03802(7) \\
0.08 & 1264 & 0.1144(6)   & 0.5352(23)  & 0.5439(26)     & 0.7093(69)     & 0.7031(53)  & 0.04328(6) \\
0.09 & 1300 & 0.1222(7)   & 0.5686(17)  & 0.5774(24)     & 0.7478(54)     & 0.7449(54)  & 0.04826(7) \\
0.10 & 1300 & 0.1302(6)   & 0.6033(19)  & 0.6116(23)     & 0.7886(65)     & 0.7866(61)  & 0.05319(6) \\
0.12 & 2500 & 0.1442(4)   & 0.6694(12)  & 0.6760(13)     & 0.8556(43)     & 0.8517(43)  & 0.06265(4) \\
0.14 & 2600 & 0.1565(3)   & 0.7384(11)  & 0.7460(13)     & 0.9321(38)     & 0.9317(39)  & 0.07181(4) \\
0.16 & 3524 & 0.1676(2)   &0.8056(8)     & 0.8142(9)        & 1.0032(29)     & 1.0029(26)  & 0.08059(3)  
\end{tabular}
\end{ruledtabular}
\end{table}

\begin{table}
\caption{Results of the spectra on $V=18^3 \times 24$ with $t_{\min}=16$.}
\label{tab:2}
\squeezetable
\begin{ruledtabular}
\begin{tabular}{lrllllll}
$m_f$ & $N_{\rm{trj}}$ & $F_\pi$ & $M_\pi$ & $M_{SC}$ & $M_{\rho(PV)}$ & $M_{\rho(VT)}$ & $\langle {\bar \psi} \psi \rangle$ \\ 
\hline
0.04 & 712 & 0.0823(5) & 0.3421(29) & 0.3445(33) & 0.4901(96) & 0.4842(107) & 0.02295(3) \\
0.05 & 752 & 0.0910(6) & 0.3886(15) & 0.3908(19) & 0.5323(84) & 0.5248(79)   & 0.02818(3) \\
0.06 & 904 & 0.0999(5) & 0.4317(15) & 0.4351(17) & 0.5900(63) & 0.5943(52)   & 0.03334(2) \\
0.07 & 1064 & 0.1090(5) & 0.4734(10) & 0.4777(13) & 0.6436(63) & 0.6398(64) & 0.03849(3) \\
0.08 & 844 & 0.1170(5) & 0.5144(10) & 0.5181(12) & 0.6782(49) & 0.6776(56)   & 0.04354(4) \\
0.10 & 876 & 0.1315(4) & 0.5948(11) & 0.5993(12) & 0.7729(65) & 0.7698(56)   & 0.05334(3)
\end{tabular}
\end{ruledtabular}
\end{table}

\begin{table}
\caption{Results of the spectra on $V=24^3 \times 32$ with $t_{\min}=22$.}
\label{tab:3}
\squeezetable
\begin{ruledtabular}
\begin{tabular}{lrllllll}
$m_f$ & $N_{\rm{trj}}$ & $F_\pi$ & $M_\pi$ & $M_{SC}$ & $M_{\rho(PV)}$ & $M_{\rho(VT)}$ & $\langle {\bar \psi} \psi \rangle$ \\
\hline
0.02 & 744 & 0.0566(8) & 0.2330(25) & 0.2367(37) & 0.3508(117) & 0.3461(111) & 0.01188(3) \\ 
0.03 & 728 & 0.0715(4) & 0.2832(14) & 0.2851(14) & 0.4044(96)   & 0.4018(101) & 0.01746(2) \\
0.04 & 864 & 0.0823(2) & 0.3353(7)   & 0.3382(7)    & 0.4678(57)   & 0.4693(57)   & 0.02290(1) \\
0.05 & 752 & 0.0918(5) & 0.3826(10) & 0.3851(11) & 0.5274(54)   & 0.5228(53)   & 0.02825(2) \\
0.06 & 1848 & 0.1012(3) & 0.4295(6)   & 0.4327(6)    & 0.5742(58)   & 0.5826(55) & 0.03345(2)\\
0.07 & 864 & 0.1088(3) & 0.4731(6)   & 0.4767(7)    & 0.6288(74)   & 0.6345(75)   & 0.03853(1) \\
0.08 & 816 & 0.1173(3) & 0.5145(8)   & 0.5187(8)    & 0.6783(81)   & 0.6795(71)   & 0.04357(2) \\
0.10 & 948 & 0.1315(3) & 0.5940(5)   & 0.5987(6)    & 0.7790(65)   & 0.7760(68)   & 0.05334(1)
\end{tabular}
\end{ruledtabular}
\end{table}

\begin{table}
\caption{Results of the spectra on $V=30^3 \times 40$ with $t_{\min}=28$.}
\label{tab:4}
\squeezetable
\begin{ruledtabular}
\begin{tabular}{lrllllll}
$m_f$ & $N_{\rm{trj}}$ & $F_\pi$ & $M_\pi$ & $M_{SC}$ & $M_{\rho(PV)}$ & $M_{\rho(VT)}$ & $\langle {\bar \psi} \psi \rangle$ \\
\hline
0.02 & 996 & 0.0578(2) & 0.2227(9)  & 0.2245(10) & 0.3225(75)   & 0.3221(65)    & 0.01189(1) \\
0.03 & 1044 & 0.0709(3) & 0.2801(7)  & 0.2818(8)   & 0.3967(87)   & 0.3940(69)  & 0.01744(1) \\
0.04 & 1060 & 0.0826(2) & 0.3354(4)  & 0.3377(4)   & 0.4718(110) & 0.4730(99)  & 0.02294(1) \\
0.05 & 1108 & 0.0918(2) & 0.3834(5)  & 0.3859(5)   & 0.5317(92)   & 0.5302(80)  & 0.02822(1) \\
0.06 & 788 & 0.1012(3) & 0.4304(4)  & 0.4332(4)   & 0.5853(132) & 0.5887(123)  & 0.03344(1) \\
0.07 & 732 & 0.1098(2) & 0.4735(4)  & 0.4769(4)   & 0.6349(141) & 0.6329(120)  & 0.03855(1) 
\end{tabular}
\end{ruledtabular}
\end{table}

\begin{table}
\caption{Results of the spectra on $V=36^3 \times 48$ with $t_{\min}=32$.}
\label{tab:5}
\squeezetable
\begin{ruledtabular}
\begin{tabular}{lrllllll}
$m_f$ & $N_{\rm{trj}}$ &  $F_\pi$ & $M_\pi$  & $M_{SC}$ & $M_{\rho(PV)}$ & $M_{\rho(VT)}$ & $\langle {\bar \psi} \psi \rangle$\\
\hline
0.015 & 1004 & 0.0512(3) & 0.1883(5) & 0.1900(7) & 0.2825(107) & 0.2904(71) & 0.00906(1) \\
0.02 & 808   & 0.0579(2) & 0.2216(7) & 0.2229(8) & 0.3202(60)    & 0.3232(61)   & 0.01189(1) \\
0.03 &  996  & 0.0704(2) & 0.2801(5) & 0.2818(6) & 0.3964(86)    & 0.3935(71)   & 0.01745(1)
\end{tabular}
\end{ruledtabular}
\end{table}
%

\section{Simulations of $N_f=4$ QCD}
\label{app:nf4}
In this appendix, we show the results of simulations for 
$N_f=4$ QCD. 
We take $\beta = 3.7$, and $(L,T)=(12, 18), (16, 24)$ and $(20,30)$. 
For each lattice size, we carry 
out simulations for various input $m_f$, and resultant values
of $F_\pi$, $M_\pi$, $M_{SC}$ and $\langle\bar{\psi} \psi\rangle$
for each parameter, 
together with the number of trajectories $N_{\rm{trj}}$ 
used for the measurements after thermalzation,
are summarized in Tables~\ref{tab:Nf4L12T18},
~\ref{tab:Nf4L16T24} and ~\ref{tab:Nf4L20T30}. 
\begin{table}
   \caption{Results of the spectra for $N_f=4$ on $V=12^3 \times 18$.}
   \label{tab:Nf4L12T18}
\squeezetable
   \begin{minipage}[t]{.7\textwidth}
   \begin{ruledtabular}
   \begin{tabular}{lrllll}
   $m_f$ & $N_{\rm trj}$ & $F_\pi$ & $M_\pi$ & $M_{SC}$ & $\langle \bar{\psi} \psi \rangle$\\
   \hline
   0.01& 500 & 0.0858(22) & 0.2373(30) & 0.3022(79) & 0.01471(48) \\
   0.02& 500 & 0.1141(11) & 0.3016(26) & 0.3502(52) & 0.02332(28) \\
   0.03& 500 & 0.12900(69)& 0.3706(11) & 0.4087(28) & 0.03056(15) \\
   0.04& 500 & 0.13842(67) & 0.4283(14) & 0.4705(21) & 0.03683(16) \\
   0.05& 500 & 0.14893(59) & 0.4778(10) & 0.5173(27) & 0.04291(29) \\
   \end{tabular}
   \end{ruledtabular}
   \end{minipage}
\end{table}
\begin{table}
   \caption{Results of the spectra for $N_f=4$ on $V=16^3 \times 24$.}
   \label{tab:Nf4L16T24}
\squeezetable
   \begin{minipage}[t]{.7\textwidth}
   \begin{ruledtabular}
   \begin{tabular}{lrllll}
   $m_f$ & $N_{\rm trj}$ & $F_\pi$ & $M_\pi$ & $M_{SC}$ & $\langle \bar{\psi} \psi \rangle$\\
  \hline
   0.005& 500 & 0.08195(76) & 0.16356(87) & 0.2235(65) & 0.01102(21) \\
   0.01& 500 & 0.10258(78) & 0.21390(75) & 0.2763(27) & 0.016413(8) \\
   0.02& 500 & 0.11980(71) & 0.2996(11) & 0.3484(16) & 0.02415(12) \\
   0.03& 500 & 0.13059(59) & 0.36739(75) & 0.4099(20) & 0.03060(11) \\
   0.04& 500 & 0.14124(31) & 0.42524(60) & 0.4659(11) & 0.03697(11) \\
   \end{tabular}
   \end{ruledtabular}
   \end{minipage}
\end{table}
\begin{table}
   \caption{Results of the spectra for $N_f=4$ on $V=20^3 \times 30$.}
   \label{tab:Nf4L20T30}
\squeezetable
   \begin{minipage}[t]{.7\textwidth}
   \begin{ruledtabular}
   \begin{tabular}{lrllll}
   $m_f$ & $N_{\rm trj}$ & $F_\pi$ & $M_\pi$ & $M_{SC}$  & $\langle \bar{\psi} \psi \rangle$\\
   \hline
   0.01& 180 & 0.10443(40) & 0.20966(68) & 0.2611(29) & 0.01637(5) \\
   0.02& 380 & 0.11955(31) & 0.29873(83) & 0.3413(12) & 0.02396(5) \\
   0.03& 350 & 0.13144(33)& 0.36699(89) & 0.4082(13) & 0.030789(36) \\
   0.04& 200 & 0.14160(24) & 0.42579(86) & 0.4663(13) & 0.037317(39) \\
   \end{tabular}
   \end{ruledtabular}
   \end{minipage}
\end{table}

\begin{figure}[!h]
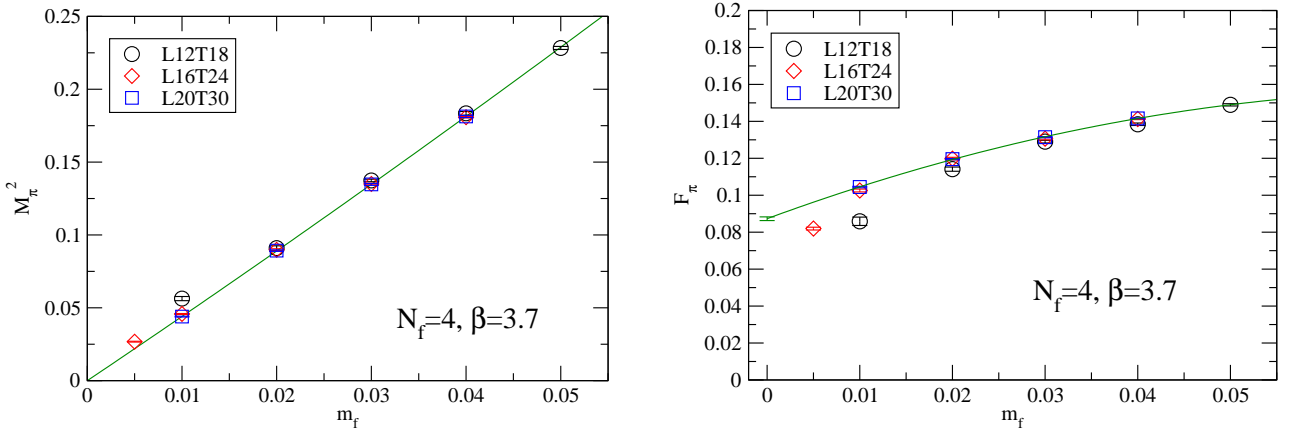
 
\makebox[.49\textwidth][r]{\includegraphics[width=.45\textwidth]{nf4_figures/mpi2_B3.7.eps}}
\makebox[.49\textwidth][r]{\includegraphics[width=.45\textwidth]{nf4_figures/fpi_B3.7.eps}}
\caption{
 $M_\pi^2$ (left panel) and $F_\pi$ as functions of $m_f$ for $N_f=4$
 QCD with $\beta=3.7$. Curves in each figure are obtained by fitting
 $c_1 m_f + c_2 m_f^2$ and $c_3 + c_4 m_f + c_5 m_f^2$ to the
 largest-volume data of $M_\pi^2$ and $F_\pi$, respectively.
}
\label{fig:nf4all}
\end{figure}

\begin{figure}[!h] 
\makebox[.49\textwidth][r]{\includegraphics[width=.45\textwidth]{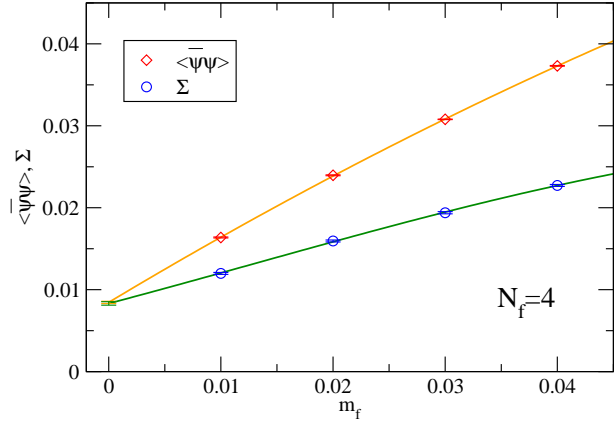}}
\caption{
 $\langle \bar{\psi} \psi \rangle$ as a function of $m_f$ for $N_f=4$
 QCD with $\beta=3.7$, $L=20$, $T=30$. The data points indicated by
 $\Sigma$ are calculated through $F_\pi^2 M_\pi^2 /(4 m_f)$ with the
 data of $M_\pi$ and $F_\pi$ at each $m_f$. Curves in the figure are the
 results of quadratic fits. 
}
\label{fig:nf4cond}
\end{figure}

In Fig.~\ref{fig:nf4all}, 
we plot $M_\pi^2$ (left panel) and $F_\pi$ as functions of $m_f$. 
Curves in each figure are obtained by fitting $c_1 m_f + c_2 m_f^2$ and $c_3 + c_4 m_f + c_5 m_f^2$ 
to the largest-volume (i.e., $L=20$, $T=30$) data of $M_\pi^2$ and $F_\pi$, respectively. 
Fit results are $c_1 = 4.36(3)$, $c_2 = 4.3(1.0)$, $c_3 = 0.0873(10)$, $c_4 = 1.846(87)$ and $c_5 = -12.3(1.6)$.
These plots show typical behavior of a theory which is in the S$\chi$SB
phase, namely, $M_\pi^2$ is well fitted by a linear function (plus small
quadratic correction) of $m_f$, and $F_\pi$ clearly has a non-zero value
in the chiral limit. As further confirmation, we also plot data of
$\langle \bar{\psi} \psi \rangle$ obtained from the largest-volume
simulation as a function of $m_f$ in Fig.~\ref{fig:nf4cond}. In the
figure, we also plotted $F_\pi^2 M_\pi^2/(4 m_f)$ ($\equiv \Sigma$)
which are calculated from the data of $M_\pi$ and $F_\pi$ at each
$m_f$. Curves in the figure are the results of quadratic fits, and
resultant values in the chiral limit are $0.00845(14)$ and $0.00832(21)$
for $\langle \bar{\psi} \psi \rangle$ and $\Sigma$, respectively. 

To estimate the amount of systematic error of chiral extrapolation, we
did the same analysis that were done in Sec.~\ref{subsec:log}. In the
plot, we show the result of ChPT extrapolation which is matched to the
quadratic fit result at $m_f=0.01$, the smallest mass we simulate. The
value of $F_\pi$ in the chiral 
limit ($F$) obtained by this procedure is $F = 0.0730$, while the
quadratic fit gives $F=0.0873(10)$. We should note here that there
is no visible chiral-log behavior in our data in the range of $0.01 \le
m_f \le 0.04$, therefore the estimate of the amount of chiral-log effect
in the chiral limit given here should be understood as the maximum
possible.  
\begin{figure}[!h] 
\makebox[.49\textwidth][r]{\includegraphics[width=.45\textwidth]{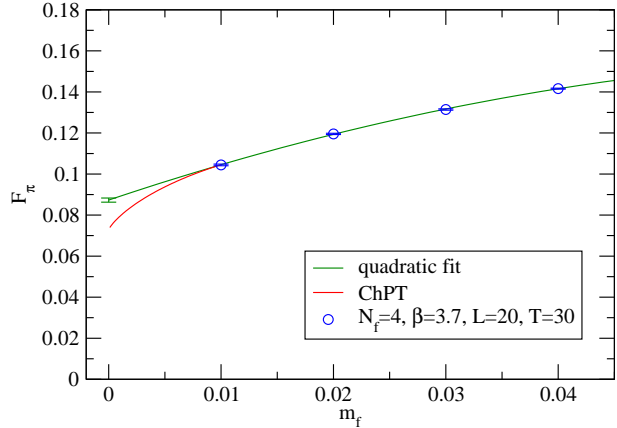}}
\caption{
Quadratic fit to the largest volume data of $F_\pi$ and ChPT extrapolation.
The value in the chiral limit by ChPT is $F_\pi = 0.0730$, while the quadratic fit gives $F_\pi=0.0873(10)$.}
\label{fig:mf4msc}
\end{figure}
We should also mention here the values of chiral expansion parameter 
$\mathcal{X}$. By using $F=0.0873$, 
the expansion parameter at $m_f^{\rm min }=0.01$ is estimated as
$\mathcal{X}\simeq0.3$ while the one at  $m_f^{\rm max}=0.04$ is
$\mathcal{X}\simeq1.2$. This confirms the consistency of using ChPT
analysis and thus, we conclude that $N_f=4$ QCD is in the S$\chi$SB phase. 

\begin{figure}[!h] 
\makebox[.49\textwidth][r]{\includegraphics[width=.45\textwidth]{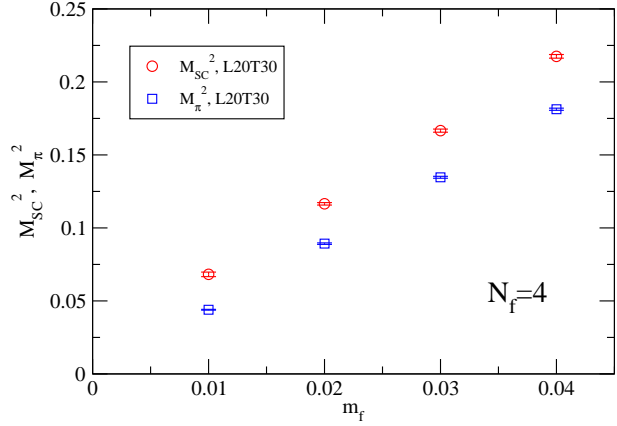}}
\caption{
Comparison of $M_{SC}^2$ and $M_\pi^2$ as a function of $m_f$ for  $N_f=4$ QCD with $\beta=3.7$. 
}
\label{fig:nf4msc}
\end{figure}

In Fig.~\ref{fig:nf4msc}, we plot the values of $M_{SC}^2$ together
with values of $M_\pi^2$ for each $m_f$. This comparison shows the
amount of flavor-symmetry-breaking effect in our simulation for $N_f=4$ QCD
with $\beta=3.7$. 

\begin{figure}[!h]
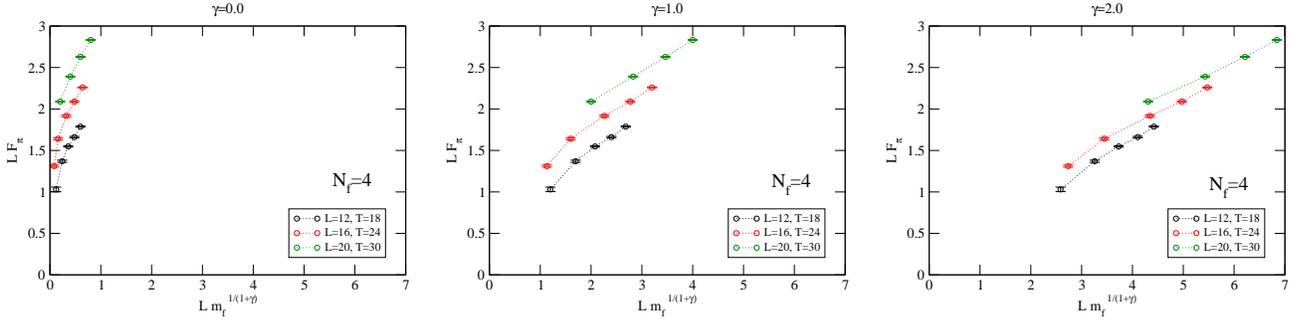
 
\makebox[.32\textwidth][r]{\includegraphics[width=.30\textwidth]{nf4_figures/fpi_B37_g00.eps}}
\makebox[.32\textwidth][r]{\includegraphics[width=.30\textwidth]{nf4_figures/fpi_B37_g10.eps}}
\makebox[.32\textwidth][r]{\includegraphics[width=.30\textwidth]{nf4_figures/fpi_B37_g20.eps}}
\caption{
 Finite size hyperscaling test of $F_\pi$ in $N_f=4$ QCD. 
 Input values of $\gamma$ are, from left to right panels, $\gamma = 0.0, 1.0$ and $2.0$, respectively.
}
\label{fig:nf4fpihs}
\end{figure}

Finally, we show the finite-size hyperscaling test for $N_f=4$ QCD 
by using the data of $F_\pi$ obtained here. In Fig.~\ref{fig:nf4fpihs}, 
we show the finite-size hyperscaling plot for input values of $\gamma= 0.0, 1.0$ and $2.0$. 
%
As we expect, the data show no alignment in the range of $0 \leq \gamma \leq 2$. 
This should be regarded as a typical property of QCD-like theory, and contrasted to the case of $N_f=8$.

\section{Estimate of chiral log corrections}
\label{app:log}

\begin{figure}[!h]
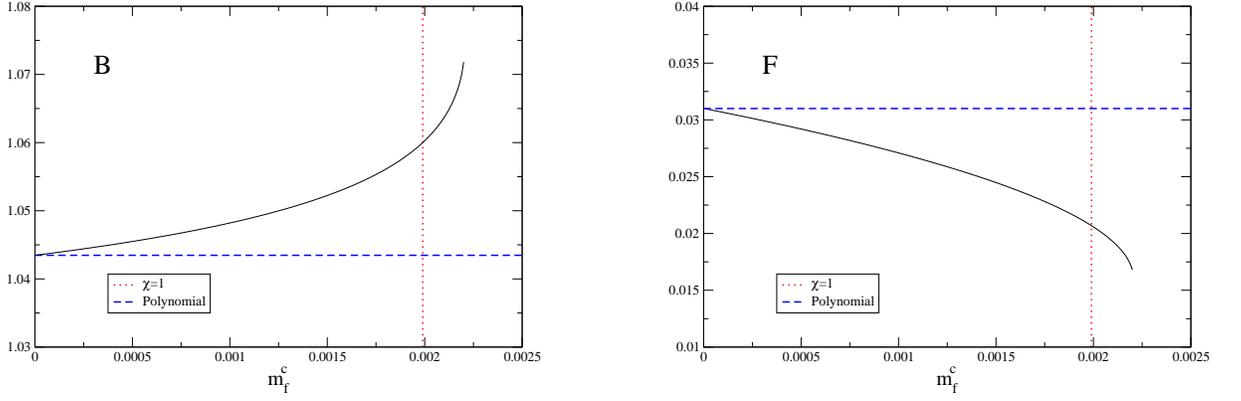
 
\makebox[.49\textwidth][r]{\includegraphics[scale=.3]{nf8_figures/B_log.eps}}
\makebox[.49\textwidth][r]{\includegraphics[scale=.3]{nf8_figures/F_log.eps}}
\caption{
Low energy constants $B$ (left) and $F$ (right) estimated by matching
 NLO ChPT formulae at the matching point $m_f^c$.
The dashed and dotted lines denote the polynomial fit result
and $m_f^c$ at ${\cal X}=1$, respectively.
}
\label{fig:log_B_F}
\end{figure}

\begin{figure}[!h] 
\makebox[.49\textwidth][r]{\includegraphics[scale=.3]{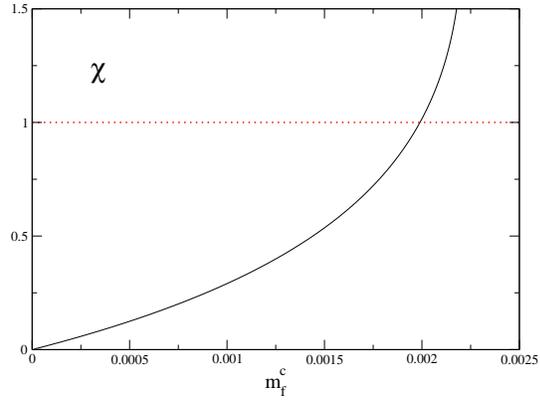}}
\caption{
${\cal X}$ estimated by matching
 NLO ChPT formulae at the matching point $m_f^c$.
 The dotted line denotes ${\cal X}=1$.
}
\label{fig:log_X}
\end{figure}

In this appendix 
we estimate the effect of the chiral log correction for the $F$ 
and $\langle \overline{\psi} \psi \rangle$ in the chiral limit.
Since both the NLO ChPT formulae Eqs.~(\ref{eq:chpt_mpi}) and 
(\ref{eq:chpt_fpi}) contain the $B$ and $F$,
to match the polynomial fit to the NLO ChPT formula 
at the matching point $m_f^c$, 
we need to solve the matching conditions using
Eqs.~(\ref{eq:chpt_mpi}), (\ref{eq:chpt_fpi}), 
and their derivatives at $m_f^c$, simultaneously.
The $m_f^c$ dependence for the $B$ and $F$ are plotted
in Fig.~\ref{fig:log_B_F}.
For comparison
the polynomial fit results are shown by the dashed lines.
The value of the ${\cal X}$ in Eq.(\ref{eq:X}) evaluated using
the obtained $F$ at each $m_f^c$ is presented in Fig.~\ref{fig:log_X}.
A reasonable value of ${\cal X} \sim 1$ is obtained only 
in much smaller $m_f$ region 
than the $m_f$ used in our simulation.
At $m_f^c = 0.00199$, 
we obtain ${\cal X}=1$ and $F=0.207$.

\begin{figure}[!h] 
\makebox[.49\textwidth][r]{\includegraphics[scale=.3]{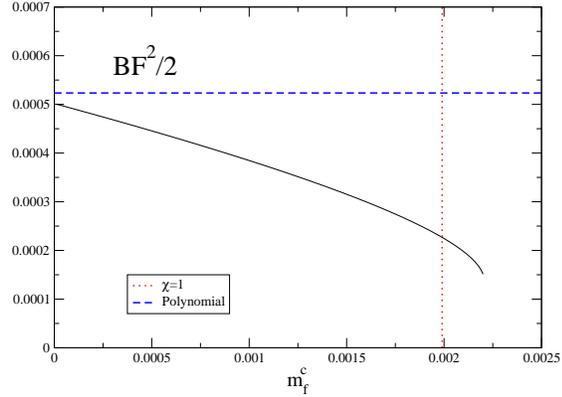}}
\caption{
 $\Sigma|_{m_f\to 0}$ estimated by matching NLO ChPT formulae at the
 matching point $m_f^c$.
The dashed and dotted lines denote the polynomial fit result
and $m_f^c$ at ${\cal X}=1$, respectively.
}
\label{fig:log_GMOR}
\end{figure}

To study the chiral log correction on the chiral condensate,
the value in the chiral limit is also estimated
from the $B$ and $F$ at each $m_f^c$ 
using the GMOR relation, 
$\displaystyle{\left. \langle \overline{\psi} \psi \rangle\right|_{m_f\to 0} = BF^2/2}$
presented in Fig.~\ref{fig:log_GMOR}.
When one chooses the matching point where ${\cal X}=1$,
one obtains roughly a half value of the polynomial fit result
of the direct measurement.

Similar results are obtained from different analyses,
such as using the expansion parameter $2 M_\pi^2/(4\pi F_\pi)^2$ 
instead of $x$ in the NLO formulae
Eqs.~(\ref{eq:chpt_mpi}) and (\ref{eq:chpt_fpi}),
or using $\langle \overline{\psi} \psi \rangle$ 
instead of $M_\pi^2/m_f$ or $F_\pi$ for matching.
Therefore we use only the result in the first case to estimate
the chiral log corrections in Sec.~\ref{subsec:log}.
%

\section{$P(\gamma)$ analysis}
\label{app:pgamma}
In this appendix in order to perform the analysis that does not assume the
functional form of fitting in FSHS,
we consider $P(\gamma)$ defined in Ref.~\cite{Aoki:2012eq}.

To quantify the ``alignment'' we introduce an evaluation function 
$P(\gamma)$ for an observable $p$ as follows. 
Suppose $\xi^j$ is a data point of the measured observable $p$ at 
$x_j = L_j \cdot m_j^{1/(1+\gamma)}$ and $\delta\xi^j$ is the error of $\xi^j$. 
$j$ labels distinction of parameters $L$ and $m_f$. 
Let $K$ be a subset of data points $\{(x_k, \xi_k)\}$ from which 
we construct a function $f^{(K)}(x)$ which represents the subset of data. 
Then, the evaluation function is defined as 
\begin{equation}
  \label{eq:p}
  P(\gamma) =
  \frac{1}{\mathcal{N}}
  \sum_{L}
  \sum_{j \not\in{K_L}}
  \frac{\left|\xi^{j} - f^{(K_L)}(x_j)\right|^2}
       {\left|\delta \xi^{j}\right|^2}, 
\end{equation}
where $L$ runs through all the lattice sizes we have,
the sum over $j$ is taken for a set of data points which do not 
belong to $K_L$ which includes all the data obtained on the lattice 
with size $L$.
$\mathcal{N}$ denotes the total number of summation. 
Here, we choose for the function $f^{(K_L)}$ a linear interpolation 
of the data points of the fixed lattice size $L$ for simplicity, 
which should be a good approximation of $\xi$ for large $x$. 

This evaluation function takes a smaller value when the data points are 
more closely collapsed to the line $f^{(K_L)}$ and thus provides 
a measure of the alignment. 
$P(\gamma)$ varies as the choice of parameter $\gamma$ and should show 
a minimum at a certain value of $\gamma$ when the optimal alignment of 
data is achieved. 
We take it as the optimal value of $\gamma$. 

We then estimate the uncertainty of the optimal $\gamma$ 
by properly taking account of the statistical fluctuation of 
$\xi_i$ as well as its effect to the line $f^{(K_L)}$. 
For this purpose, we employ the parametric bootstrap method,
in which the data point is simulated by a random sample 
generated by Gaussian distribution 
with the mean $\xi^j$ and the standard deviation $\delta\xi^j$. 
The distribution of $\gamma$ is thus obtained for a large number 
of these samples, from which the variance of $\gamma$ is estimated. 
The systematic error associated with the interpolation will be 
estimated by choosing different functional form with linear or 
quadratic splines as will be discussed subsequently. 

\begin{figure}[!h] 
\makebox[.45\textwidth][r]{\includegraphics[scale=.40,trim=0 0 0 0,clip=true]{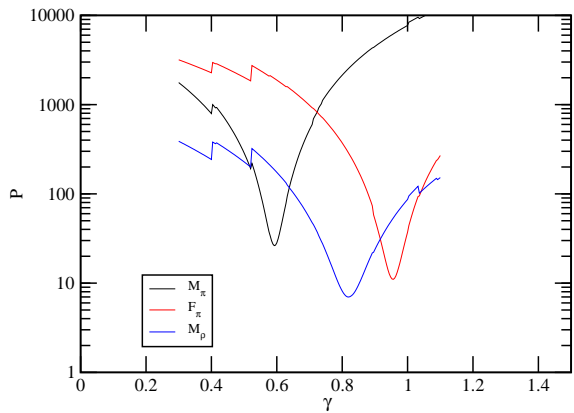}}
\caption{
$P(\gamma)$ in $M_\pi$ $F_\pi$  and $M_\rho$ in $N_f=8$ QCD.
}
\label{fig:pgamma}
\end{figure}

We use the data as the overlapped region 
sandwiched between $m_f=0.015$ on $L=36$ 
and $m_f=0.03$ on $L=36$.
Fig.~\ref{fig:pgamma} is the result of $P(\gamma)$ for all $M_\pi$, $F_\pi$ and $M_\rho$
and there are minima of $P(\gamma)$.
The $\gamma$ value at the minimum of $P(\gamma)$  is written on Table~\ref{tab:pgamma}.
%
\begin{table}[!h]
\caption{Optimal $\gamma$ to make $P(\gamma)$ minimum with statistical error.
}
\label{tab:pgamma}
\begin{minipage}[t]{.7\textwidth}
\begin{ruledtabular}
\begin{tabular}{cddd}
&
\multicolumn{1}{c}{$M_\pi$} &
\multicolumn{1}{c}{$F_\pi$} &
\multicolumn{1}{c}{$M_\rho$} \\
\hline
$\gamma$   &  0.593(2) & 0.955(4) &  0.820(20) \\
\end{tabular}
\end{ruledtabular}
\end{minipage}
\end{table}
%


\bibliographystyle{apsrev4-1}
\bibliography{nf8_main}

\end{document}